\documentclass[iop,numberedappendix,appendixfloats]{emulateapj} 
\submitted{}

\usepackage{amsmath}
\usepackage{natbib} 
\usepackage[backref,breaklinks,colorlinks,citecolor=blue]{hyperref} 
\usepackage[flushleft]{threeparttable} 
\usepackage{xcolor} 
\usepackage{adjustbox}

\newcommand\numberthis{\addtocounter{equation}{1}\tag{\theequation}}

\begin{document}

\begin{abstract}
We study the relationship between dense gas and star formation in the Antennae galaxies by comparing ALMA observations of dense gas tracers (HCN, HCO$^+$, and HNC $\mathrm{J}=1-0$) to the total infrared luminosity ($\mathrm{L_{TIR}}$) calculated using data from the \textit{Herschel} Space Observatory and the \textit{Spitzer} Space Telescope. We compare the luminosities of our SFR and gas tracers using aperture photometry and employing two methods for defining apertures. We taper the ALMA dataset to match the resolution of our $\mathrm{L_{TIR}}$ maps and present new detections of dense gas emission from complexes in the overlap and western arm regions. Using OVRO CO $\mathrm{J}=1-0$ data, we compare with the total molecular gas content, $\mathrm{M(H_2)_{tot}}$, and calculate star formation efficiencies and dense gas mass fractions for these different regions. We derive HCN, HCO$^+$ and HNC upper limits for  apertures where emission was not significantly detected, as we expect emission from dense gas should be present in most star-forming regions. The Antennae extends the linear $\mathrm{L_{TIR}-L_{HCN}}$ relationship found in previous studies. The $\mathrm{L_{TIR}-L_{HCN}}$ ratio varies by up to a factor of $\sim$10 across different regions of the Antennae implying variations in the star formation efficiency of dense gas, with the nuclei, NGC 4038 and NGC 4039, showing the lowest SFE$_\mathrm{dense}$ (0.44 and 0.70 $\times10^{-8}$ yr$^{-1}$). The nuclei also exhibit the highest dense gas fractions ($\sim 9.1\%$ and $\sim7.9\%$).
\end{abstract}


\title{kiloparsec-scale Variations in the Star Formation Efficiency of Dense Gas: the Antennae Galaxies (NGC 4038/39)}
\author{Ashley Bemis, Christine Wilson}
\affil{McMaster University}

\section{Introduction}

The Antennae galaxies are the nearest pair of merging galaxies (22 Mpc, \citealt{Schweizer:2008}) and are rich in star formation (e.g. \citealt{Whitmore:1999}), gas (e.g. \citealt{Wilson:2000,Wilson:2003}), and dust (e.g. \citealt{Klaas:2010}).  The rarity of wet, major mergers (gas-rich galaxies with a mass ratio $\le$ 3) makes the Antennae a particularly unique environment for studying star formation in interactions. Recent simulations suggest the Antennae is $\sim40$ Myr after its second pass \citep{Karl:2010}, placing it at an intermediate stage in the Toomre sequence. Thus, the Antennae contains multiple generations of stars from merger-induced starburst behavior. The two nuclei exhibit post-starburst populations $\sim$65 Myr old \citep{Mengel:2005}, and even younger starburst populations ($\sim3-10$ Myr) are concentrated in the overlap region and western arm (e.g. \citealt{Mengel:2001, Mengel:2005, Whitmore:2010,Whitmore:2014}). Furthermore, different regions within the Antennae exhibit varying degrees of current ($\le 100$ Myr, \citealt{Brandl:2009}) star formation, with the overlap region of the Antennae (see Fig. \ref{fig:data}) experiencing a particularly violent episode ( Star Formation Rate, SFR $>4$ M$_\odot$ yr$^{-1}$, \citealt{Brandl:2009,Klaas:2010}; this work). 

Major mergers are a testbed for the extreme star formation ongoing at high-z, and show fundamental differences in their star formation properties compared with normal star-forming disk galaxies (e.g. \citealt{Daddi:2010,Tacconi:2018}).  Futhermore, star formation occurs primarily in the densest regions within Giant Molecular Clouds (GMCs, $\mathrm{n(H_2)}>10^4$ cm$^{-3}$ , \citealt{Lada:1991a,Lada:1991b}). The HCN $\mathrm{J}=1-0$ transition has a critical density of $\mathrm{n_{crit}}\sim10^5$ cm$^{-3}$, while the CO $\mathrm{J}=1-0$ has $\mathrm{n_{crit}}\sim10^2$ cm$^{-3}$. Thus, it is essential to observe molecules such as HCN to constrain the properties of the directly star-forming gas. 


Extragalactic studies often use observations of the total infrared luminosity ($\mathrm{L_{IR}}$) and HCN J$=1-0$ molecular luminosity ($\mathrm{L_{HCN}}$) in galaxies to study star formation and dense gas. This has largely been motivated by the seminal work of \cite{Gao:2004a,Gao:2004b}, who  found a tight and linear relationship between the global values of $\mathrm{L_{IR}}$ and $\mathrm{L_{HCN}}$ in a sample of 65 galaxies. Their observations were of unresolved systems, thus comparing the Total Infrared (TIR) and HCN luminosities spanning $\mathrm{L_{IR}}\sim10^9-10^{12}$ L$_\odot$. This sample included normal star-forming galaxies as well as more extreme Luminous and Ultraluminous Infrared Galaxies (LIRGs/ULIRGs), suggesting a direct scaling between the SFR and dense molecular gas content across galaxy types. Other recent studies show that this linear relationship also extends to the scales of individual, massive clumps in the Milky Way and nearby galaxies (e.g. \citealt{Wu:2005,Wu:2010,Bigiel:2015,Chen:2015}), spanning nearly 10 orders of magnitude in luminosity. These observations have motivated density-threshold models of star formation \citep{Lada:2012}, which assume that star formation begins once the gas reaches a threshold density ($\mathrm{n(H_2)}=10^4$ cm$^{-3}$). These models predict a constant Star Formation Efficiency of dense gas (SFE$_\mathrm{dense}$) that should span all regimes of star formation.


A number of recent studies target the $\mathrm{L_{TIR}-L_\mathrm{HCN}}$ relationship on a variety of scales, down to several hundred parsecs \citep{Kepley:2014,Bigiel:2016,Gallagher:2018}. These studies fit well within the scatter of the original \cite{Gao:2004a,Gao:2004b} relationship, extending it down to lower luminosities. Some have also revealed variations in the $\mathrm{L_{IR}}$ and $\mathrm{L_{HCN}}$ relationship at $\sim$kpc scales (e.g. M51 from \citealt{Chen:2015}; \citealt{Usero:2015}). \cite{Usero:2015} study $\sim$kpc scales across the disks of normal star-forming galaxies and find a sublinear power-law index ($\sim0.5$) for their sample of galaxies.  Furthermore, evidence exists that (U)LIRGs may turn off the linear portion of the $\mathrm{L_{IR}-L_{HCN}}$ sequence \citep{Gracia-Carpio:2008}, suggesting variations at the high luminosity end as well.


A separate class of star formation models that can, to some degree, better explain the variations of the $\mathrm{L_{IR}-L_{HCN}}$ relationship are turbulence-regulated density threshold models \citep{KrumholzMcKee:2005, Padoan:2011}. These models predict the variation of probability density profiles (PDFs) as a function of turbulence, and show that turbulence acts as a star formation inhibitor and subsequently increases the threshold density of gas required for star formation. Observational evidence of a correlation between stellar mass density and lower $\mathrm{L}_\mathrm{TIR}/\mathrm{L}_\mathrm{HCN}$ in disk galaxies supports the idea that stellar feedback, in the form of turbulence, etc., can inhibit star formation per unit dense gas mass \citep{Bigiel:2016}. Interestingly, there have been observations of increases in the dense gas fraction (often traced by $\mathrm{L}^{'}_\mathrm{HCN}/\mathrm{L}^{'}_\mathrm{CO}$) in the central regions of disk galaxies, where the star formation efficiency of dense gas (traced by $\mathrm{L}_\mathrm{TIR}/\mathrm{L}^{'}_\mathrm{HCN}$) appears lowest and stellar density appears highest. The Central Molecular Zone (CMZ) of the Milky Way is the closest example of an environment with low SFE$_\mathrm{dense}$ and high dense gas fractions (e.g. \citealt{Kauffmann:2017a,Kauffmann:2017b}) compared to the solar neighborhood. There are a number of possible mechanisms that can explain this, with turbulence being the favored mechanism so far \citep{Federrath:2012, Kruijssen:2014,Rathborne:2014}. \cite{Federrath:2012} compare the expectations of six different star formation with Magnetohydrodynamic (MHD) simulations that vary four fundamental parameters: virial parameter, sonic mach number, turbulent forcing parameter, and Alfven mach number. They find turbulence is the primary regulator of the SFR, and produce star formation efficiencies of the total gas (SFE) that agree well with observations ($1-10\%$).



High-resolution Atacama Large Millimeter/submillimeter Array (ALMA) observations have revealed HCN, HCO$^+$, and HNC $\mathrm{J}=1-0$ emission throughout star-forming regions in the Antennae \citep{Schirm:2016}. Assuming these transitions trace $\mathrm{n(H_2)}>10^4$ cm$^{-3}$, this suggests there is an abundance of dense gas throughout this system. Futhermore, there are interesting variations in the molecular luminosities of these dense gas tracers, suggesting differences in dense gas properties across the system. \cite{Schirm:2016} found evidence for variations of the dense gas fraction across the Antennae, evidenced by higher HCN-to-CO luminosity ratios in the two nuclei when compared to the overlap region (see Fig. \ref{fig:data}). \cite{Bigiel:2015} find that the $\mathrm{L_{IR}-L_{HCN}}$ relationship in the brightest regions of the Antennae galaxies is consistent with the linear relationship revealed by \cite{Gao:2004a,Gao:2004b}, but their sensitivity limits miss a large portion of the star-forming regions in the system (e.g. the western arm and fainter regions in the overlap region). In this paper, we attempt to understand the variations of this relationship in the context of the Antennae galaxies by assessing the variations of the physical properties with $\mathrm{L}_\mathrm{TIR}-\mathrm{L}^{'}_\mathrm{HCN(1-0)}$ at subgalactic scales.


In \S \ref{sec:data}, we present the ALMA, \textit{Herschel}, and \textit{Spitzer} data used in our study along with the total infrared luminosity calibrations. In \S \ref{sec:analysis}, we describe our aperture photometry analyses. In \S \ref{sec:results}, we present the luminosity fit results and compare to previous work. In \S \ref{sec:discussion}, we discuss the variation we see in $\mathrm{SFE_{dense}}$ across the Antennae and explore potential explanations for these variations. The analysis and results of this study are summarized in \S \ref{sec:conclusion}. Molecular and infrared luminosity uncertainties are discussed in more detail in Appendix \ref{app:uncertainties}. A comparison between total infrared luminosity calbrations from \cite{Galametz:2013} is presented in Appendix \ref{app:ltir}.

\section{Data} \label{sec:data}

\begin{figure*}[tb]
    \centering
    \includegraphics[width=0.7\textwidth]{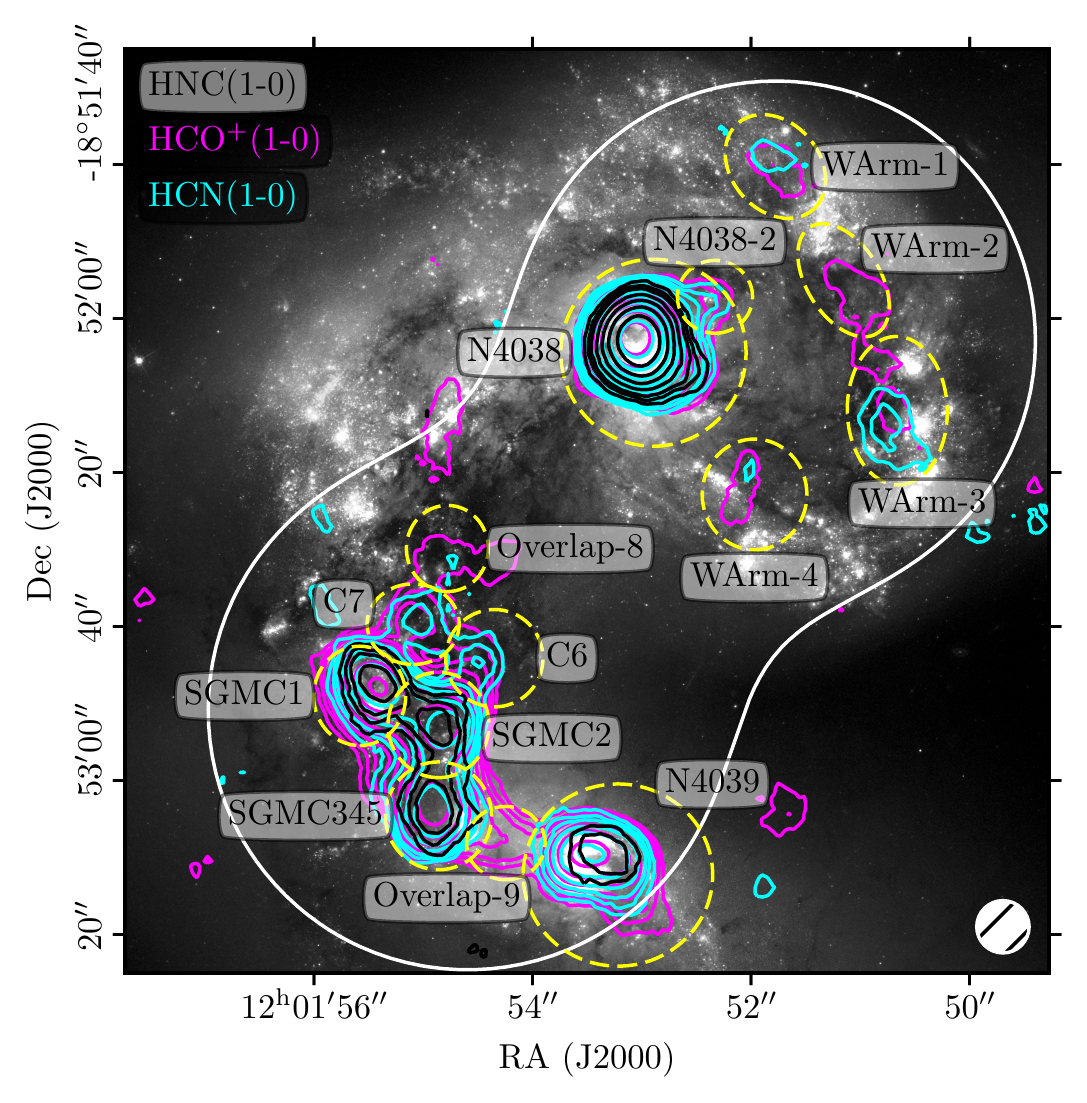}
    \caption{Magenta, cyan, and black contours (levels = $0.06\times$[4,  6,  8,  12,  17, 24,  34,  49,  70,  100] Jy beam$^{-1}$ km s$^{-1}$) of the ALMA HCO$^+$, HCN, and HNC J$=1-0$ transitions, respectively, overlaid on top of a black and white composite image (435 nm, 550 nm, and 658 nm) from \textit{HST}. The elliptical apertures are outlined with yellow, dashed curves and labeled according to Tables \ref{tab:ell_ap} and \ref{tab:ell_lumin}; the 50$\%$ ALMA primary beam sensitivity is shown as the solid, white curve. The smoothed beam size ($6.8''$) of the ALMA data is shown in the lower right.}
    \label{fig:data}
\end{figure*}

We use \textit{Herschel}, \textit{Spitzer}, and ALMA data in our study to compare star formation traced by infrared emission to dense gas traced by high critical-density molecular transitions, HCN, HCO$^+$, and HNC $\mathrm{J}=1-0$ (see Figure \ref{fig:data}). We also use CO $\mathrm{J}=1-0$ data from the Owens Valley Radio Observatory (OVRO, \citealt{Wilson:2003}) as our bulk molecular gas tracer, and we note that the OVRO data may be missing $\sim20\%$ of the CO $\mathrm{J}=1-0$ flux \citep{Schirm:2016}, likely a diffuse component of the gas, due to the limited range of $u-v$ coverage. Our resolution is limited by the \textit{Herschel} data ($5.5''$ at 70 $\mu$m, and  $6.8''$ at 100 $\mu$m), and thus our analysis is performed at these resolutions.

\subsection{ALMA Data}

Details on the observations of the ALMA data are available in \cite{Schirm:2016}. The original reduction scripts were used to apply calibrations to the raw data using the appropriate Common Astronomy Software Applications (\textsc{CASA}) version (\textsc{CASA} 4.2.0, \citealt{McMullin:2007}). The ALMA data were then cleaned and imaged in  in \textsc{CASA} 4.7.2. We cleaned using a velocity resolution of $\Delta \mathrm{v_{opt}} = 5.2\ \mathrm{km\ s}^{-1}$ at the rest frequency of each transition over an optical velocity range of 1000-2000 km s$^{-1}$. We tapered the data to the Full-Width Half-Maximum (FWHM) of the \textit{Herschel} 70 $\mu$m Point Spread Function using a Briggs weighting of $0.5$ while cleaning. The largest angular scale of the ALMA observations is $\sim 17''$\footnote{ALMA Cycle 1 Proposer's Guide} ($\sim 1.8$ kpc). The tapered data reach a root mean square noise level (rms) of $\sigma=1.2\ \mathrm{mJy\ beam}^{-1}$. When working at the $100\ \mu$m resolution, we further smooth the tapered cube to $6.8''$.

We create moment zero maps of the molecular lines using \textsc{CASA}'s immoments command. This produces a two-dimensional image of the integrated intensity with units of Jy beam$^{-1}$ km s$^{-1}$. We require that all pixels going into the final moment map be greater than $2\sigma$, where $\sigma=1.2$ mJy beam$^{-1}$ in the $5.5''$ maps and $\sigma=1.4$ mJy beam$^{-1}$ in the smoothed 6.8$''$ maps. We then convert to molecular luminosities ($\mathrm{L'_{mol}}$) using the following equation \citep{Wilson:2008}

\begin{align*}
\frac{\mathrm{L'_{mol}}}{\textrm{K km s}^{-1}\ \textrm{pc}^2} &= 3.2546 \times 10^7 \left( \frac{\mathrm{S_{ap}}}{\textrm{Jy km s}^{-1}} \right)\times \\
	&\qquad\left( \frac{\mathrm{D_L}}{\textrm{Mpc}} \right)^{2}\left( \frac{\nu_0}{\textrm{GHz}} \right)^{-2}(1 + \mathrm{z})^{-1} \numberthis \label{eq:lumin}
\end{align*}

\noindent where $\mathrm{S_{ap}}$ is the flux measured in an aperture in $\textrm{Jy km s}^{-1}$. This gives molecular luminosity in units of $\textrm{K km s}^{-1}\textrm{ pc}^2$. We use a redshift of $\mathrm{z}=0.005477$. Details on the uncertainty estimates are given in \S \ref{sec:analysis} and  Appendix \ref{app:mol_lumin}. 

\subsection{Infrared Data and Total Infrared Luminosities} \label{sec:ltir}

We obtain user-provided data products of the 70, 100, 160, and 250 $\mu$m maps from the \textit{Herschel} \citep{Pilbratt:2010} Science Archive. Details on the observations and reduction of the 70, 100, and 160 $\mu$m (Photodetector Array Camera and Spectrometer) PACS \citep{Poglitsch:2010} data are available in \cite{Klaas:2010} and reach resolutions of $5.5''$, $6.8''$, and $11.3''$, respectively. The Spectral and Photometric Imaging Receiver (SPIRE) 250 $\mu$m map ($18.1''$ resolution, \citealt{Griffin:2010}) was obtained as part of the Very Nearby Galaxies Survey and details on the observations and calibrations can be found in \cite{Bendo:2012:VNGS}. We also retrieve user-provided \textit{Spitzer} \citep{Werner:2004:Spitzer} 24 $\mu$m Multiband Imaging Photometer (MIPS) data (\citealt{Rieke:2004:MIPS}, $6.0''$ resolution) from the \textit{Spitzer} Heritage Archive. These data were reprocessed by \cite{Bendo:2012} to provide ancillary data for the \textit{Herschel}-SPIRE Local Galaxies Guaranteed Time Programs.

We use several calibrations from \cite{Galametz:2013} to estimate L$_\mathrm{TIR}$, which is defined in that paper to be:
\begin{align*}
\mathrm{L_{TIR}} = \int_{3\ \mu\mathrm{m}}^{1100\ \mu\mathrm{m}} \mathrm{L}_\nu \mathrm{d}\nu.
\end{align*}
\noindent \cite{Galametz:2013} derive calibrations of L$_\mathrm{TIR}$ using a combination of \textit{Herschel} and \textit{Spitzer} data from $8-250\ \mu \mathrm{m}$ as an alternative to fitting the dust spectral energy distribution (SED). They have provided monochromatic calibrations (e.g. 70 $\mu$m), as well as multi-band calibrations (e.g. 24+70+100 $\mu$m). We compare several of these calibrations for the Antennae in Appendix \ref{app:ltir} and show the ratio maps for these calibrations in Figure \ref{fig:ltir250} at the 250 $\mu$m resolution ($18.1''$).

In this paper, we use the monochromatic 70 $\mu$m (5.5'' $\sim$ 590 pc) and the multi-band $24+70+100\ \mu$m  (6.8'' $\sim$ 725 pc) calibrations  to estimate L$_\mathrm{TIR}$ across the Antennae. The 70 $\mu$m calibration is the highest-resolution \textit{Herschel} band and brackets the warm-dust (30-60 K) Spectral Energy Distribution (SED) peak ($\sim100\ \mu$m). For multi-band calibrations, \cite{Galametz:2013} recommend that any L$_\mathrm{TIR}$ estimate using less than 4-5 bands should include the 100 $\mu$m flux or a combination of the $70+160$ bands, which should lead to L$_\mathrm{TIR}$ predictions reliable within 25\% ($\le50\%$ for monochromatic calibrations). Additionally, \cite{Galametz:2013} note that including the 24 $\mu$m flux improves calibrations of L$_\mathrm{TIR}$ for galaxies with higher 70/100 color, i.e. strongly star-forming environments. The overlap region is known to be vigorously star-forming, which could cause the 70 $\mu$m flux to underestimate L$_\mathrm{TIR}$. Therefore, we include the $24+70+100\ \mu$m calibration as a check for this.  Overall, we find our L$_\mathrm{TIR}(70)$ estimates agree well with the L$_\mathrm{TIR}(24+70+100)$ estimates. The L$_\mathrm{TIR}(70)$ estimate for SGMC345 (the combination of SGMCs 3, 4, and 5 from \citealt{Wilson:2000}) is only $\sim3\%$ lower  than the L$_\mathrm{TIR}(24+70+100)$ estimate and agrees within uncertainties. 

To estimate L$_\mathrm{TIR}$ using multiple IR bands, we converted the \textit{Herschel} and \textit{Spitzer} maps to the same units and resolution (i.e. to the FWHM of the beam size of the band with the lowest resolution). The \textit{Spitzer} MIPS and \textit{Herschel} SPIRE data were converted to units of Jy pixel$^{-1}$ from MJy sr$^{-1}$ and Jy beam$^{-1}$, respectively (the \textit{Herschel} PACS data were already in units of Jy pixel$^{-1}$). Each dataset was then convolved to a common resolution using the \cite{Aniano:2011} kernels. The \citet{Galametz:2013} calibrations require infrared measurements be in solar luminosity units ($\text{L}_\odot$). We convert the \textit{Herschel} infrared maps from Jansky units to solar luminosities using the following equation
\begin{align*} \label{eq:lir_watts}
\frac{\nu \mathrm{L}_\nu}{\mathrm{L_\odot}} = 3.1256\times10^{2} \left(\frac{\mathrm{d_L}}{\mathrm{Mpc}}\right)^2 \left( \frac{\nu}{\mathrm{GHz}} \right) \left(\frac{ \mathrm{S}_\nu}{\mathrm{Jy}}\right). \numberthis
\end{align*}

\noindent The background is then estimated and subtracted from each map. Once the data are formatted properly, we apply the corresponding \cite{Galametz:2013} calibrations to create L$_\mathrm{TIR}$ maps. We calculate absolute uncertainties on the L$_\mathrm{TIR}$ calibrations (see Appendix \ref{app:ltir_uncer} for details) and find they are much lower than the calibration uncertainties quoted above (25$\%$ uncertainties on the L$_\mathrm{TIR}(24+70+100)$ measurements and $\sim50\%$ uncertainties on the L$_\mathrm{TIR}$(70) measurements).

\section{Aperture Analysis} \label{sec:analysis}

We compare the emission of our SFR and gas tracers across different regions of the Antennae using aperture photometry. We use two approaches to defining apertures. In our first method, we identify clumps of emission using \textsc{cprops} \citep{Rosolowsky:2006} in each of the dense gas data-cubes; we then manually define elliptical apertures (Table \ref{tab:ell_ap}) to encompass infrared and integrated intensity dense-gas emission of individual ``clumps'' or complexes\footnote{There are multiple clumps of dense gas emission along most lines-of-sight, but in an aperture-photometry analysis we sum over all of this emission.}. We vary the radii and position angles of the apertures to encompass potentially-associated emission of the IR and dense-gas tracers. In our second method, we perform a ``pixel-by-pixel" analysis by dividing the maps into hexagonal grids that are sampled by the FWHM of the beam (i.e. the incircle diameter of each hexagon is $6.8''$). The hexagons are fixed in size across the map (edge = $3.9''$; inspired by a similar method employed by \citealt{Leroy:2016}). The elliptical aperture method allows us to contrast the behavior of individual regions, while the hexagonal method eliminates selection bias that can be introduced in the manual-aperture method. Therefore, the hexagonal aperture method provides more robust data for trend-fitting, and the emphasis of the elliptical aperture analysis is region-by-region comparisons.

\begin{table}[tb]
\begin{threeparttable}
\caption{\sc Elliptical Apertures}
\label{tab:ell_ap}
\centering
\setlength{\tabcolsep}{4pt}
\begin{tabular}{lllrrr}
\hline\hline
\multicolumn{1}{c}{Source}	&	\multicolumn{1}{c}{RA}	&	\multicolumn{1}{c}{Dec}	&	\multicolumn{1}{c}{EW}	&	\multicolumn{1}{c}{NS}	& \multicolumn{1}{c}{Area} \\
	&	\multicolumn{1}{c}{(J2000)}	&	\multicolumn{1}{c}{(J2000)}	&	\multicolumn{1}{c}{($^{\prime\prime}$)}	&	\multicolumn{1}{c}{($^{\prime\prime}$)}	&	\multicolumn{1}{c}{(kpc$^2$)} \\
	\hline
NGC4038 & $12:01:52.895$ & $-18:52:04.46$ & 24.1 & 24.3 & 5.23 \\
NGC4039 & $12:01:53.22$ & $-18:53:12.29$ & 24.6 & 23.7 & 5.22 \\
NGC4038-2 & $12:01:52.332$ & $-18:51:57.2$ & 9.8 & 9.5 & 0.84 \\
WArm-1 & $12:01:51.779$ & $-18:51:40.27$ & 11.4 & 14.9 & 1.52 \\
WArm-2 & $12:01:51.156$ & $-18:51:55.04$ & 9.5 & 16.3 & 1.39 \\
WArm-3 & $12:01:50.664$ & $-18:52:11.98$ & 13.0 & 19.3 & 2.26 \\
WArm-4 & $12:01:51.97$ & $-18:52:22.86$ & 13.6 & 14.4 & 1.76 \\
SGMC1 & $12:01:55.583$ & $-18:52:49.02$ & 12.0 & 12.9 & 1.38 \\
SGMC2 & $12:01:54.862$ & $-18:52:52.8$ & 13.1 & 13.7 & 1.60 \\
SGMC345 & $12:01:54.862$ & $-18:53:04.6$ & 13.8 & 14.0 & 1.73 \\
Schirm-C6 & $12:01:54.351$ & $-18:52:44.13$ & 12.6 & 12.7 & 1.44 \\
Schirm-C7 & $12:01:55.094$ & $-18:52:39.71$ & 12.0 & 10.5 & 1.12 \\
Overlap-8 & $12:01:54.781$ & $-18:52:29.88$ & 10.6 & 11.2 & 1.05 \\
Overlap-9 & $12:01:54.246$ & $-18:53:08.14$ & 10.2 & 9.5 & 0.86 \\

\hline
\end{tabular}
\begin{tablenotes}
\item {\sc Note. --} The center coordinates and angular extent of the elliptical apertures. The axes are oriented East-West (EW) and North-South (NS) except for WArm-1, WArm-2, and WArm-3 where the position angles (PA) are 42.2$^\circ$, 32.3$^\circ$, and 1.3$^\circ$ east of north.
\end{tablenotes}
\end{threeparttable}
\end{table}

We perform the luminosity-luminosity fits using the Bayesian linear regression code \textsc{linmix} \citep{Kelly:2007},  which incorporates uncertainties in both x- and y-directions. The \textsc{linmix} routine assumes a linear relationship of the form $\mathrm{log(L_{TIR}) = m \times log(L_{dense})+ log(b)}$, where m is the slope, and b is the y-intercept. The \textsc{linmix} code also allows us to incorporate upper limits into our fits, therefore we also include upper limits of the molecular luminosities in the fits. We estimate the significance of each correlation by calculating the Spearman rank coefficients of the datasets for each fitted relationship. The one- and two-sigma uncertainties on the fits are estimated via Markov-Chain Monte-Carlo (MCMC), and we take the median values of these iterations as our fit parameters. We compare our results with those of \cite{Gao:2004a,Gao:2004b} and \citet{Liu:2015} and also perform fits of the Antennae data combined with datasets from these studies. We also compare with measurements of $\mathrm{L_{TIR}}$ and $\mathrm{L_{HCN}}$ of the CMZ  \citep{Stephens:2016} since we observe similarities between luminosity ratios of the CMZ and the two nuclei (see \S \ref{sec:cmz_similarities}). The HCN luminosity for the CMZ is derived from the Mopra CMZ 3mm survey, covering a $2.5^\circ\times0.5^\circ$ area centered on $\mathit{l}=0.5^\circ$, $\mathit{b}=0.0^\circ$ \citep{Jones:2012}, and the conversion to luminosity assumes a distance of 8.34$\pm$0.16 kpc \citep{Reid:2014}. The infrared luminosity of the CMZ is estimated using a combination of 12, 25, 60, and 100 $\mu$m  Infrared Astronomical Satellite (IRAS) fluxes and the calibration from \cite{Sanders:1996}. We note that this region of the Milky Way is very crowded, and these luminosities are likely upper limits and may include emission from other sources along the line of sight. We assume uncertainties of $\sim30\%$ on $\mathrm{L_{TIR}}$ and $\mathrm{L_{HCN}}$ of the CMZ since these are also the uncertainties prescribed by \citet{Liu:2015} to the galaxies in their sample.

To determine upper limits of the Antennae luminosities, we first estimate the contribution of noise into each moment zero map. We approach this differently than applying a simple rms noise limit due to the large physical extent of the apertures  (i.e. larger than the GMCs in our beam), and the large velocity range over which our moment maps are created. The moment zero maps are created with a two-sigma cutoff, such that no emission below two-sigma is allowed into the map. Since the noise follows a Gaussian distribution, only $\sim$2$\%$ of the noise should remain above this two-sigma cutoff; choosing a cutoff at two-sigma allows us to eliminate the majority of the noise from the moment zero map without sacrificing a significant amount of real emission. However, because $\sim$2$\%$ of the noise remains, each aperture will contain some signal from the noise proportional to the number of pixels per aperture ($\mathrm{n_{pix}}\simeq117$ for the hexagonal apertures, and varies for the elliptical apertures) and the total number of channels in the datacube ($\mathrm{n_{chan}} = 192$). Furthermore, the noise varies with position in the map according to the response of the primary beam ($\epsilon_\mathrm{pb}$). Optimally, the base rms noise is $\sigma=1.2$ mJy beam$^{-1}$ (5.5$''$) or $\sigma=1.4$ mJy beam$^{-1}$ (6.8$''$) at an efficiency of 100$\%$, and larger as the response decreases towards the edges of the primary beam.

Therefore, the remaining Gaussian noise per aperture in the moment zero maps, $\mathrm{\sigma^{ap}_{Gauss}}$, is
\begin{align} \label{eq:ap_noise}
\mathrm{\sigma^{ap}_{Gauss} \approx 0.025 \times \frac{2\sigma}{\epsilon_{pb}} \times \Delta v \times \mathrm{n_{pix}} \times \mathrm{n_{chan}}\div ppb} 
\end{align}
\noindent where $\Delta \mathrm{v} = 5.2$ km s$^{-1}$ and ppb is the number of pixels per beam (to give units of Jy km s$^{-1}$).  We require the aperture sums from the moment zero maps of the dense gas tracers to be larger than this noise limit to be considered a detection. We set our upper limit to two times this noise limit:
\begin{align} \label{eq:limit}
\mathrm{\sigma^{ap}_{lim}} &= 2\times \mathrm{\sigma^{ap}_{Gauss}}
\end{align}

\noindent This requires at least $0.025\times192=5$ channels (per pixel per aperture) to have a signal of four-sigma to be considered a detection. Anything below $\mathrm{\sigma^{ap}_{lim}}$ is considered an upper limit, and we set the values of these apertures to $\mathrm{\sigma^{ap}_{lim}}$ and treat them as upper limits in our fitting routines. The upper limits are shown as the gray arrows in Figure \ref{fig:hex_pixels}. At the 6.8$''$ resolution, the limit per hexagonal aperture is set to be $\mathrm{\sigma^{ap}_{lim}}\approx$ 110 mJy km s$^{-1}$ at maximum primary beam efficiency (which corresponds to 0.9 mJy km s$^{-1}$ per pixel), and translates to luminosity limits of $\mathrm{log(L_{HCN})} = 5.46$, $\mathrm{log(L_{HCO^+})} = 5.46$, and $\mathrm{log(L_{HNC})} = 5.45$.

\section{Results} \label{sec:results}

Each approach to defining apertures has its own strengths: the hexagonal-grid approach allows us to optimally sample our datasets without introducing selection bias into our apertures, and the elliptical aperture analysis has the benefit of emphasizing individual source behavior. We therefore focus on the hexagonal aperture results when discussing fits, and then later focus on the results of the elliptical apertures when discussing variations in different regions of the Antennae. 

In the tapered ALMA dataset, we detect significant emission from HCN and HCO$^+$  in both the nuclei (NGC 4038 and NGC 4039), the overlap region (containing SGMCs 1-5, C6 and C7 from \citealt{Schirm:2016}, and newly-detected sources 8 and 9), and the western arm (containing WArm 1-4). HNC is detected significantly in NGC 4038 and SGMCs in the overlap region, and upper limits are derived elsewhere. HCO$^+$ is the overall brightest dense gas tracer in this dataset, and there are several regions where we detect HCO$^+$ but not HCN; this includes the``bridge" region (Overlap-8) between SGMCs 3, 4, and 5 (hereafter referred to as one source, SGMC345) and NGC 4039 (the southern nucleus). HCO$^+$ is also brightest in one of the regions we study in the western arm (WArm-2).

\begin{figure*}[th]
	\flushright
    \includegraphics[width=0.95\textwidth]{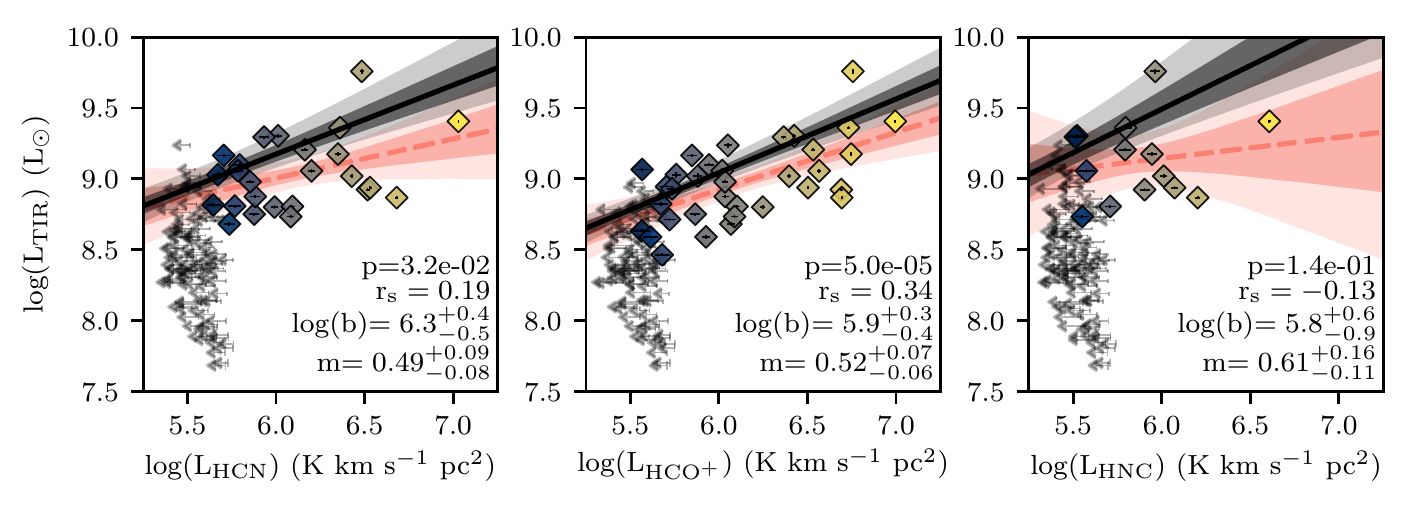}
    \flushleft
    \includegraphics[width=0.99\textwidth]{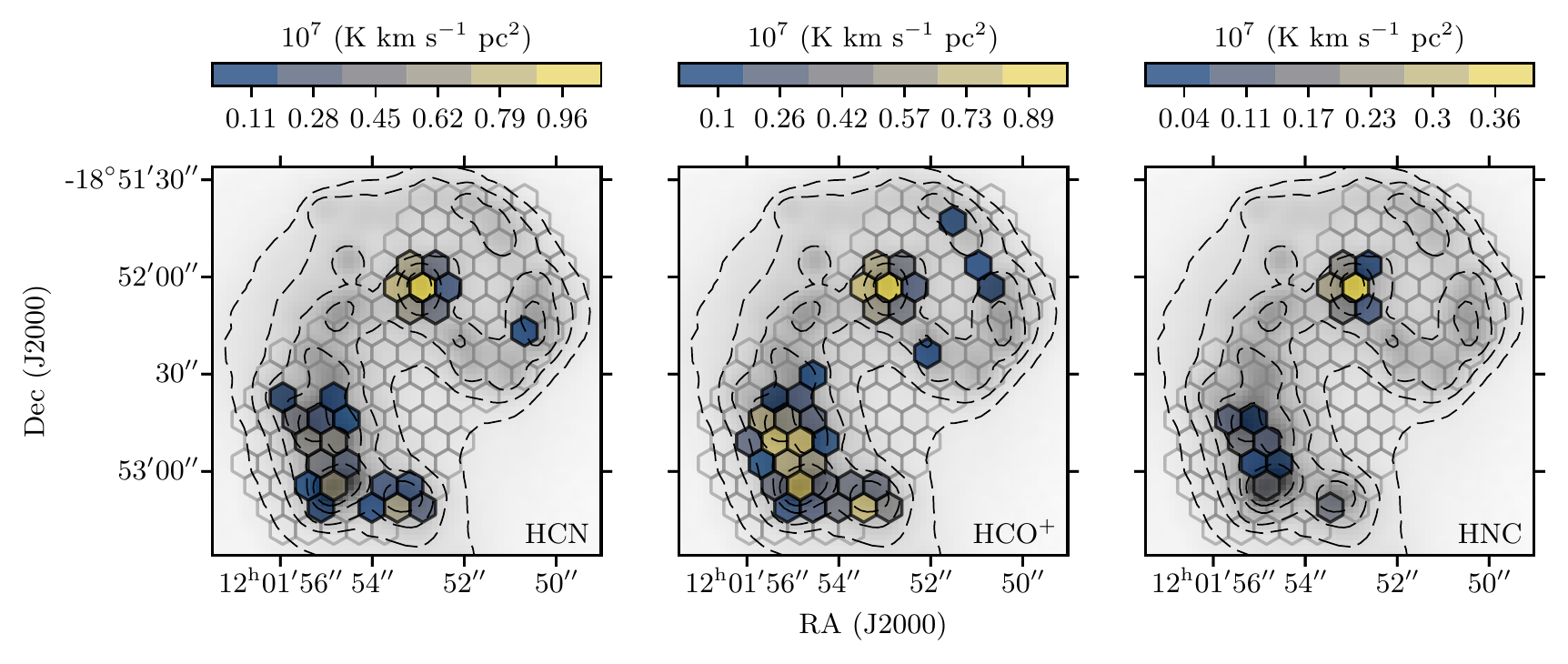}
    \caption{\label{fig:hex_pixels} \textit{Top:} From left to right, we show $\mathrm{L_{TIR}}$ vs. $\mathrm{L_{HCN}}$, $\mathrm{L_{HCO^+}}$, and $\mathrm{L_{HNC}}$. The datapoints are colorized according to the hexagonal apertures in the maps directly below; gray datapoints and open apertures are upper limits. The \textsc{linmix} fit including upper limits is shown as the solid black line. The fit without upper limits is shown for comparison as the salmon dashed line. The one-sigma (dark shaded area) and two-sigma (light shaded area) uncertainties resulting from the MCMC iterations are also shown. The resulting fits assume a linear relationship of the form $\mathrm{log(L_{TIR}) = m \times log(L_{dense})+ log(b)}$, where m is the slope, and b is the y-intercept; we show the resulting slopes (m) and y-intercepts (b) for the fits including upper limits on the plots. The absolute uncertainties are plotted on each datapoint, which are generally small for log(L$_\mathrm{TIR}$). The Spearman rank coefficients for each correlation are also shown.  \textit{Bottom:} From left to right, the hexagonal apertures are shown for HCN, HCO$^+$, and HNC overlaid on top of the $\mathrm{L_{TIR}}$ map (dashed contours and grayscale, log stretch). The hexagons are colorized according to the luminosity of dense gas emission corresponding to that aperture. The colorbar values are in units of 10$^{7}$ K km s$^{-1}$ pc$^{-2}$ (log stretch).}
\end{figure*}

\begin{figure*}[tbh]
    \flushright
    \includegraphics[width=0.99\textwidth]{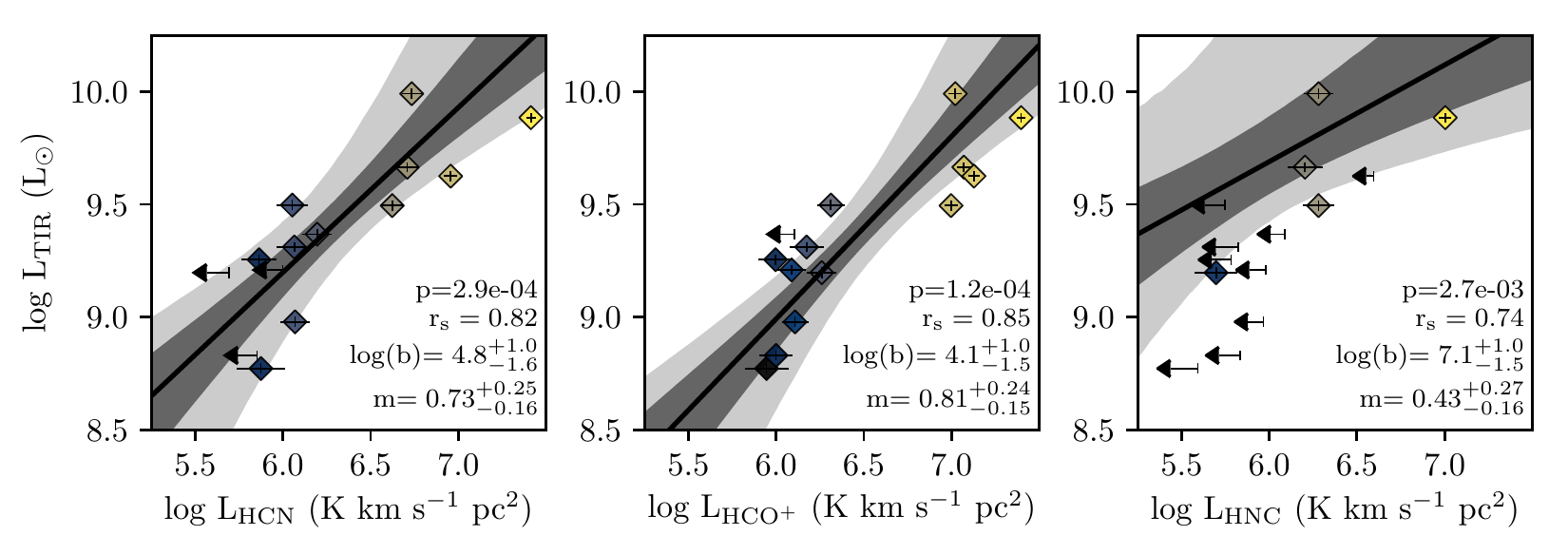}
    \flushleft
    \includegraphics[width=0.99\textwidth]{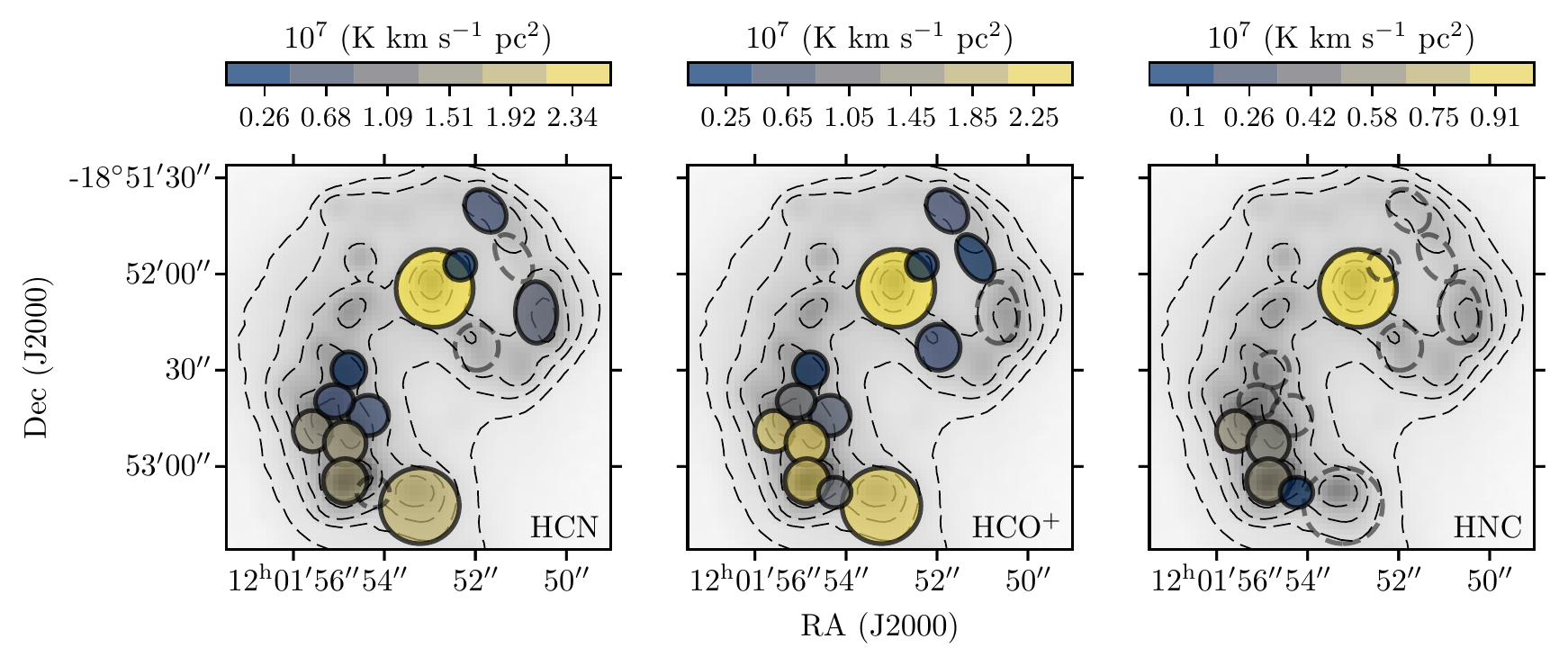}
    \caption{\label{fig:ell_results} \textit{Top:} From left to right, we show $\mathrm{L_{TIR}}$ vs. $\mathrm{L_{HCN}}$, $\mathrm{L_{HCO^+}}$, and $\mathrm{L_{HNC}}$ resulting from the elliptical aperture analysis. The datapoints are colorized according to the elliptical apertures in the maps directly below; black datapoints and open apertures are upper limits. \textit{Bottom:} From left to right, the elliptical apertures are shown for HCN, HCO$^+$, and HNC overlaid on top of the $\mathrm{L_{TIR}}$ map (dashed contours and grayscale, log stretch). We do not plot fits without upper limits.}
\end{figure*}

We plot the $\mathrm{L_{TIR}-L_{dense}}$ Antennae datapoints of the hexagonal and elliptical apertures in Figures \ref{fig:hex_pixels} and \ref{fig:ell_results}, respectively, and we list the luminosities measured within the elliptical apertures in Table \ref{tab:ell_lumin}. We overplot the \textsc{linmix} fits in grayscale (including upper limits) for both the hexagonal and elliptical aperture analyses; for comparison, we also plot fits to the hexagonal aperture luminosities without upper limits (salmon). The slopes and y-intercepts of the fits that includes upper limits are shown in each plot. We list the fits from the hexagonal apertures below. The $\mathrm{L_{TIR}-L_{dense}}$ fits present sub-linear power-law indices (i.e. $\mathrm{m}<1.0$).  The Spearman p-values indicate strong correlations ($\mathrm{p}<0.05$) between $\mathrm{L_{TIR}}$ and the dense gas molecular luminosities, except for the HNC fits which shows a weaker correlation ($\mathrm{p}\sim0.14$) likely due to the lower detection rate of this molecular line.

\begin{table*}[tb]
\centering
\begin{threeparttable}
\caption{\sc Total Infrared and Molecular Luminosities}
\label{tab:ell_lumin}
\begin{tabular}{l|c|c|c|c|c}\hline\hline
\multicolumn{1}{c}{Source}	&	\multicolumn{1}{c}{$\mathrm{L_{TIR}}$\tnote{a}}	&	\multicolumn{1}{c}{$\mathrm{L_{HCN}}$}	&	\multicolumn{1}{c}{$\mathrm{L_{HCO^+}}$}	&	\multicolumn{1}{c}{$\mathrm{L_{HNC}}$}	&	\multicolumn{1}{c}{$\mathrm{L_{CO}}$} \\[0.2em]
\multicolumn{1}{c}{}	&	\multicolumn{1}{c}{($10^9$ L$_\odot$)}	& \multicolumn{4}{c}{($10^7$ K km s$^{-1}$ pc$^{2}$)}	\\ \hline
NGC4038 & 7.67$\pm$0.39 & 2.60$\pm$0.16 & 2.50$\pm$0.15 & 1.008$\pm$0.081 & 38.4$\pm$7.9 \\
NGC4039 & 4.22$\pm$0.21 & 0.905$\pm$0.082 & 1.34$\pm$0.11 & $<$0.039 & 18.0$\pm$4.0 \\
NGC4038-2 & 0.592$\pm$0.031 & 0.075$\pm$0.024 & 0.088$\pm$0.025 & $<$0.093 & 2.92$\pm$0.91 \\
WArm-1 & 0.953$\pm$0.049 & 0.117$\pm$0.023 & 0.128$\pm$0.022 & $<$0.123 & 3.90$\pm$0.97 \\
WArm-2 & 0.676$\pm$0.035 & $<$0.100 & 0.100$\pm$0.021 & $<$0.095 & 4.2$\pm$1.1 \\
WArm-3 & 2.33$\pm$0.12 & 0.16$\pm$0.03 & $<$0.068 & $<$0.066 & 3.32$\pm$0.83 \\
WArm-4 & 1.619$\pm$0.083 & $<$0.073 & 0.123$\pm$0.023 & $<$0.070 & 2.01$\pm$0.57 \\
SGMC1 & 3.12$\pm$0.16 & 0.420$\pm$0.063 & 0.995$\pm$0.087 & 0.19$\pm$0.04 & 16.2$\pm$3.6 \\
SGMC2 & 4.62$\pm$0.24 & 0.512$\pm$0.079 & 1.18$\pm$0.11 & 0.160$\pm$0.037 & 22.7$\pm$5.0 \\
SGMC345 & 9.79$\pm$0.51 & 0.543$\pm$0.071 & 1.052$\pm$0.094 & 0.191$\pm$0.037 & 17.2$\pm$3.8 \\
Schirm-C6 & 2.05$\pm$0.11 & 0.116$\pm$0.028 & 0.149$\pm$0.034 & $<$0.061 & 4.7$\pm$1.3 \\
Schirm-C7 & 3.14$\pm$0.16 & 0.113$\pm$0.024 & 0.205$\pm$0.037 & $<$0.056 & 6.3$\pm$1.6 \\
Overlap-8 & 1.797$\pm$0.093 & 0.073$\pm$0.017 & 0.099$\pm$0.022 & $<$0.068 & 4.6$\pm$1.2 \\
Overlap-9 & 1.575$\pm$0.087 & $<$0.049 & 0.182$\pm$0.036 & 0.050$\pm$0.014 & 1.75$\pm$0.55 \\

\hline
\end{tabular}
\begin{tablenotes}
\item {\sc Note. --} Luminosities measured from the elliptical apertures listed in Table \ref{tab:ell_ap}. All values are measured at the 100 $\mu$m resolution (6.8$''$). The absolute uncertainties are shown next to each luminosity, except in the case of limits.
\item[a] This $\mathrm{L_{TIR}}$ is estimated using the \cite{Galametz:2013} calibration that combines the \textit{Spitzer} 24 and \textit{Herschel} 70 and 100 $\mu$m maps. See Table \ref{tab:ltir} for a comparison with \cite{Galametz:2013} monochromatic 70 $\mu$m $\mathrm{L_{TIR}}$ estimates.
\end{tablenotes}
\end{threeparttable}
\end{table*}

The fits from the hexagonal apertures shown in Figure \ref{fig:hex_pixels} are as follows:
\begin{align*}
\mathrm{log}(\mathrm{L_{TIR}}) &= 6.3^{+0.4}_{-0.5} + 0.49^{+0.09}_{-0.08} \mathrm{log}(\mathrm{L_{HCN}}) \numberthis \label{fit:ltirlhcnHex} \\
\mathrm{log}(\mathrm{L_{TIR}}) &= 5.9^{+0.3}_{-0.4} + 0.52^{+0.07}_{-0.06} \mathrm{log}(\mathrm{L_{HCO^+}}) \numberthis \label{fit:ltirlhcoHex} \\
\mathrm{log}(\mathrm{L_{TIR}}) &= 5.8^{+0.6}_{-0.9} + 0.61^{+0.16}_{-0.11} \mathrm{log}(\mathrm{L_{HNC}}) \numberthis \label{fit:ltirlhncHex}
\end{align*}

\begin{figure*}[tbh]
    \centering
    \label{fig:ltir_lhcn_fit_all}
    \includegraphics[width=0.99\textwidth]{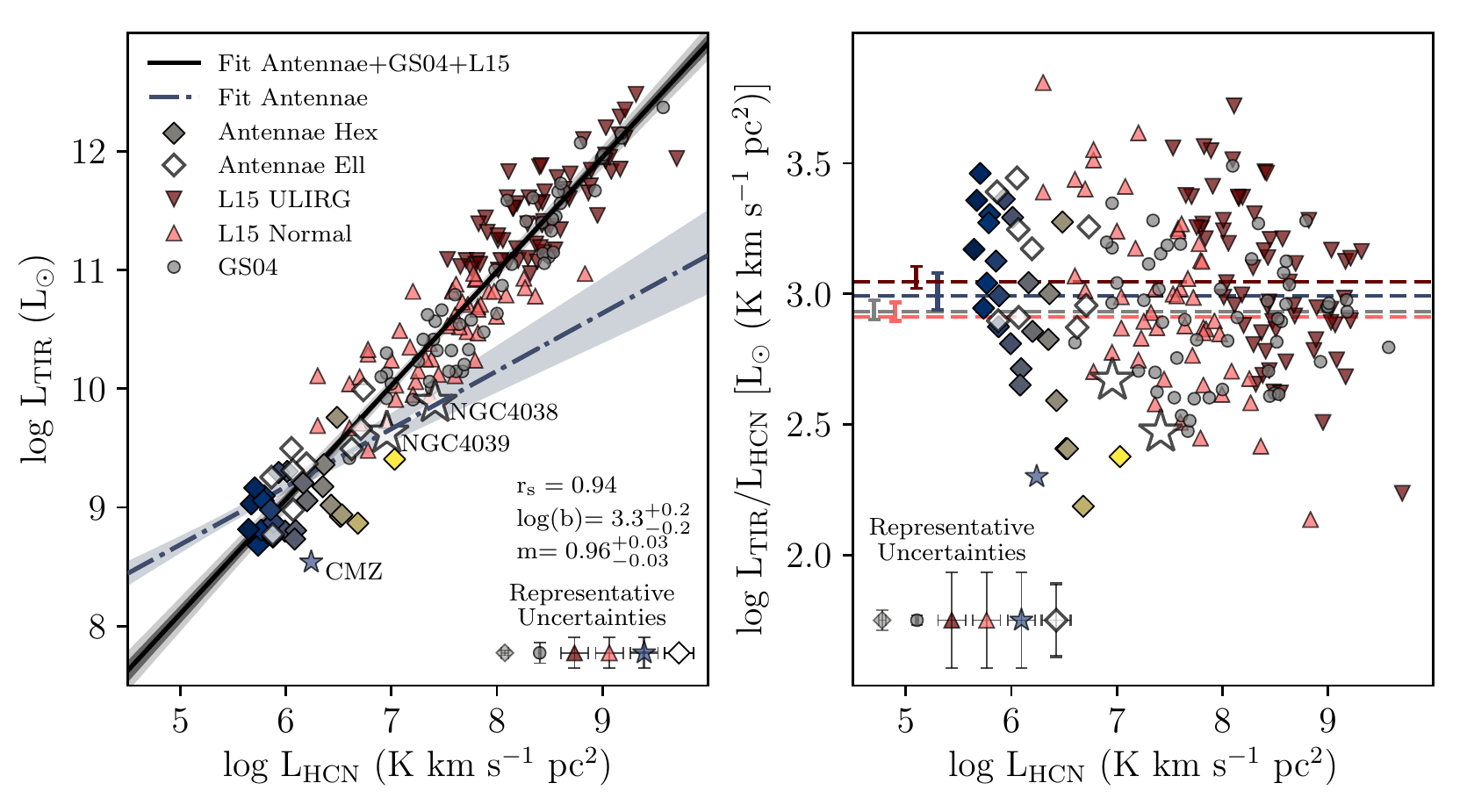}
    \caption{ {\it Left:} L$_\mathrm{TIR}$ vs. L$_\mathrm{HCN}$ datapoints of the \cite{Gao:2004a,Gao:2004b} sample (gray circles), the \cite{Liu:2015} (U)LIRGs (maroon inverted triangles), the \cite{Liu:2015} normal star-forming galaxies (salmon triangles), the Antennae hexagonal aperture luminosities (blue-yellow diamonds), and the Antennae elliptical aperture luminosities (white diamonds). We emphasize the datapoints of the two nuclei, NGC 4038 and NGC 4039, as white stars. The hexagonal datapoints are colorized according to HCN luminosity as is done in Fig. \ref{fig:hex_pixels}. We show measurements of the CMZ (as given in \citealt{Stephens:2016}) for comparison as the blue star (see text for more information). We show the fit to the \cite{Liu:2015} and Antennae hexagonal data points (black solid line), with the one- (light shade) and two-sigma (dark shade) uncertainties from the MCMC iterations. The resulting fit parameters are listed (assuming $\mathrm{log(L_{TIR}) = m \times log(L_{dense})+ log(b)}$, where m is the slope, and b is the y-intercept). The fit to the Antennae hexagonal apertures from Fig. \ref{fig:hex_pixels} is shown for comparison as the dash-dot line in blue.  {\it Right:} $\mathrm{L_{TIR}/L_{HCN}}$ vs. L$_\mathrm{HCN}$. We show the median value (dashed line, see Table \ref{tab:stats}) of the $\mathrm{L_{TIR}/L_{HCN}}$ ratio for each dataset, excluding the Antennae elliptical apertures. At the left end of the dashed lines, we show the statistical uncertainties on the median values. We show representative uncertainties in the lower right corner (left plot) and lower left corner (right plot). Upper limits are excluded. The two nuclei have the lowest $\mathrm{L_{TIR}/L_{HCN}}$ in the Antennae system, despite having the highest dense gas fractions. The CMZ is also known to have a low star formation efficiency of dense gas and very high dense gas fractions. }
\end{figure*}

\begin{figure}[th]
	\centering
	\label{fig:surf_densities}
    \includegraphics[width=0.49\textwidth]{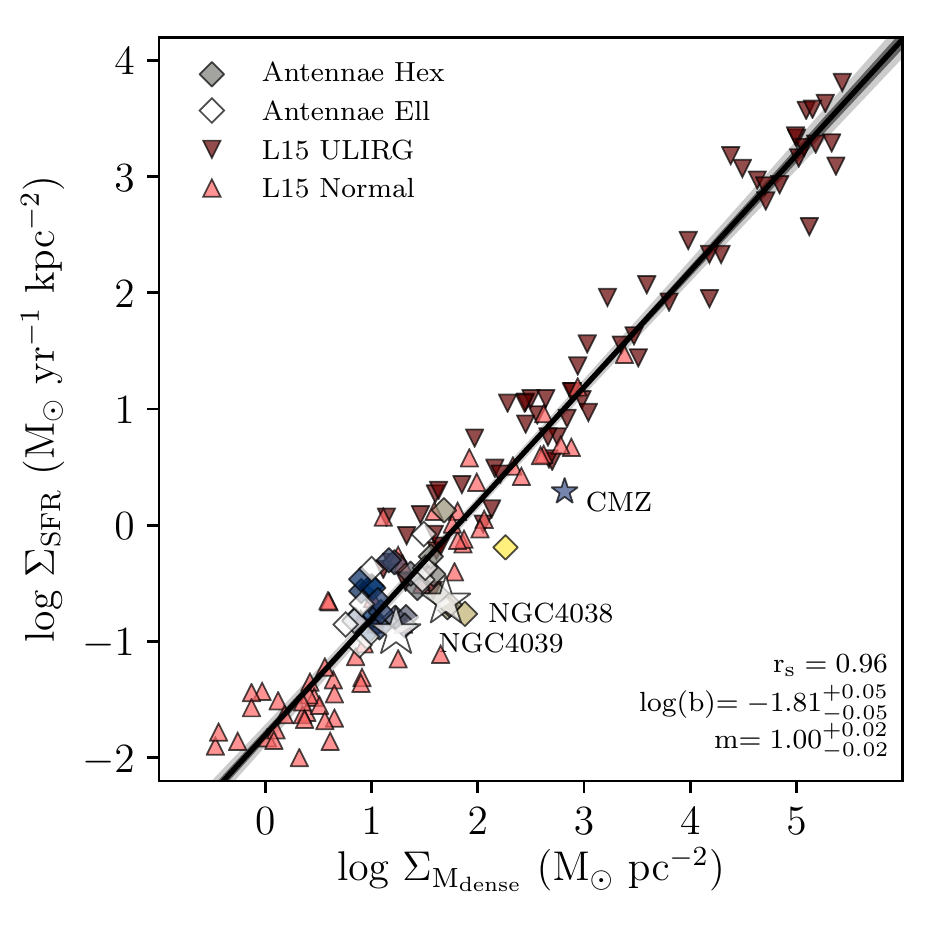}
    \caption{ $\mathrm{\Sigma_{SFR}}$ vs. $\Sigma_\mathrm{M_{dense}}$, excluding the data from  \cite{Gao:2004a,Gao:2004b}. The symbols and colors are the same as in Figure \ref{fig:ltir_lhcn_fit_all}. \cite{Liu:2015} calculate source sizes using the radio continuum. \cite{Liu:2015} calculate the dense molecular gas mass assuming an HCN conversion factor, $\alpha_\mathrm{HCN}=10$ M$_\odot$ (K km s$^{-1}$ pc$^2$). We use the SFRs that \cite{Liu:2015} calculate from IR luminosities. The two nuclei of the Antennae have surface densities comparable to those of other regions, and the locus of Antennae points falls in the overlapping regime of normal star-forming galaxies and (U)LIRGs from the \cite{Liu:2015} sample. The fit to the Antennae hexagonal datapoints and \cite{Liu:2015} datapoints yields a slope of $\mathrm{m}=1.00\pm0.02$. Upper limits are excluded.}
\end{figure}

We also fit the $\mathrm{L_{TIR}}$ and $\mathrm{L_{HCN}}$ values from the Antennae hexagonal apertures with those of the sources in \cite{Liu:2015}; \cite{Liu:2015} includes the data from the \cite{Gao:2004a,Gao:2004b} survey. This fit is presented in Figure \ref{fig:ltir_lhcn_fit_all}. The power-law index on this fit is $\mathrm{m}=0.96\pm0.03$, which is slightly sublinear. The Antennae data extend the datapoints of \cite{Gao:2004a,Gao:2004b} and \cite{Liu:2015} to lower luminosities. This agrees with the findings of \cite{Bigiel:2015}, who performed a similar analysis on the Antennae using data from the Combined Array for Research in Millimeter-wave Astronomy (CARMA). The median value of the $\mathrm{L_{TIR}/L_{HCN}}$ ratio of the Antennae hexagonal apertures (980 $\mathrm{L_\odot}$ (K km s$^{-1}$ pc$^2$)$^{-1}$) falls within the scatter of other studies (i.e. \citealt{Gao:2004a,Gao:2004b,Liu:2015}). The scatter of the Antennae data is also comparable ($\sim$0.4 dex) to these other studies (Table \ref{tab:stats}). Fitting the surface densities of the \cite{Liu:2015} and Antennae data also yields a slope $\mathrm{m=1}$ (Fig. \ref{fig:surf_densities}).

\begin{table}[tb]
\begin{threeparttable}
\caption{\sc Comparison of $\mathrm{L_{TIR}/L_{HCN}}$ Statistics}
\label{tab:stats}
\centering
\setlength{\tabcolsep}{4pt}
\begin{tabular}{l|llr|rr}
\hline\hline
    & \multicolumn{3}{c}{Non-Gaussian} &  \multicolumn{2}{|c}{Gaussian}  \\
\multicolumn{1}{c}{Dataset}	&	\multicolumn{1}{|c}{Median}	&	\multicolumn{1}{c}{Lower\tnote{a}}	&	\multicolumn{1}{c}{Upper\tnote{b}}	& \multicolumn{1}{|c}{Mean} &
\multicolumn{1}{c}{1-$\sigma$} \\
	\hline
GS04 & 850 & 440 & 620 & 950 & 550 \\
L15 Normal & 820 & 320 & 1100 & 1200 & 1100 \\
L15 ULIRGs & 1100 & 510 & 1200 & 1400 & 980 \\
Antennae (Hex.) & 980 & 610 & 990 & 1100 & 750 \\
Antennae (Ell.) & 900 & 270 & 1200 & 1300 & 780 \\

\hline
\end{tabular}
\begin{tablenotes}
\item {{\sc Note.} -- We calculate both Gaussian (mean, standard deviation) and non-Gaussian statistics (median, $16^\mathrm{th}$, and $84^\mathrm{th}$ percentiles), excluding upper limits. Quantities are in units of $\mathrm{L_\odot}$ (K km s$^{-1}$ pc$^2$)$^{-1}$. \vspace{0.3em}} 
\item[a] Distance from the median to the $16^\mathrm{th}$ percentile. 
\item[b] Distance from the median to the $84^\mathrm{th}$ percentile.
\end{tablenotes}
\end{threeparttable}
\end{table}

\begin{table*}[tb]
\centering
\begin{threeparttable}
\caption{\sc Physical Properties}\label{tab:ell_props}
\begin{tabular}{lccccccccc}\hline\hline
\multicolumn{1}{c}{Source}	&	\multicolumn{1}{c}{SFR}	&	\multicolumn{1}{c}{M(H$_2$)$_\mathrm{dense}$}	&	\multicolumn{1}{c}{M(H$_2$)}	&	\multicolumn{1}{c}{f$_\mathrm{dense}$}	&	\multicolumn{1}{c}{SFE$_\mathrm{dense}$}	&	\multicolumn{1}{c}{SFE} &   $\Sigma_\mathrm{SFR}$   &   $\Sigma_\mathrm{M{dense}}$  & $\Sigma_\mathrm{M(H_2)}$ \\
	&	\multicolumn{1}{c}{(M$_\odot$ yr$^{-1}$)}	& \multicolumn{2}{c}{($10^7$ M$_\odot$)}	&	\multicolumn{1}{c}{($\%$)}	&	\multicolumn{2}{c}{($10^{-8}$ yr$^{-1}$ )} & \multicolumn{1}{c}{(M$_\odot$ yr$^{-1}$ kpc$^{-2}$)}  & \multicolumn{2}{c}{(M$_\odot$ pc$^{-2}$)}  \\ \hline
NGC4038 & 1.1 & 26 & 270 & 9.7 & 0.44 & 0.043 & 0.22 & 50.0 & 510 \\
NGC4039 & 0.63 & 9.0 & 130 & 7.2 & 0.69 & 0.050 & 0.12 & 17.0 & 240 \\
NGC4038-2 & 0.088 & 0.75 & 20 & 3.7 & 1.2 & 0.043 & 0.11 & 9.0 & 240 \\
WArm-1 & 0.14 & 1.2 & 27 & 4.3 & 1.2 & 0.052 & 0.093 & 7.7 & 180 \\
WArm-2 & 0.10 & $<$0.71 & 29 & $<$2.4 & $>$1.4 & 0.034 & 0.073 & $<$5.1 & 210 \\
WArm-3 & 0.35 & 1.6 & 23 & 6.8 & 2.2 & 0.15 & 0.15 & 6.9 & 100 \\
WArm-4 & 0.24 & $<$1.0 & 14 & $<$7.1 & $>$2.4 & 0.17 & 0.14 & $<$5.7 & 80 \\
SGMC1 & 0.46 & 4.2 & 110 & 3.7 & 1.1 & 0.041 & 0.34 & 30.0 & 820 \\
SGMC2 & 0.69 & 5.1 & 160 & 3.2 & 1.3 & 0.043 & 0.43 & 32.0 & 990 \\
SGMC345 & 1.5 & 5.4 & 120 & 4.5 & 2.7 & 0.12 & 0.84 & 31.0 & 700 \\
Schirm-C6 & 0.30 & 1.2 & 33 & 3.5 & 2.6 & 0.092 & 0.21 & 8.1 & 230 \\
Schirm-C7 & 0.47 & 1.1 & 44 & 2.5 & 4.1 & 0.11 & 0.42 & 10.0 & 400 \\
Overlap-8 & 0.27 & 0.73 & 32 & 2.3 & 3.7 & 0.083 & 0.26 & 7.0 & 310 \\
Overlap-9 & 0.23 & $<$0.49 & 12 & $<$4.0 & $>$4.8 & 0.19 & 0.27 & $<$5.7 & 140 \\

\hline
\end{tabular}
\begin{tablenotes}
\item {\sc Note. --} Estimates of physical properties from sources contained within the elliptical apertures (see Table \ref{tab:ell_ap}). These properties are derived from the luminosities listed in Table \ref{tab:ell_lumin}. See \S \ref{sec:results} for further information about the conversion from luminosities to these values and the corresponding uncertainties.
\end{tablenotes}
\end{threeparttable}
\end{table*}

We convert the total infrared luminosity to estimates of the star formation rates using the calibration initially published in \cite{Kennicutt:1998} and updated in \cite{Kennicutt:2012} with the more recent Kroupa initial mass function and Starburst99 model \citep{Hao:2011,Murphy:2011}
\begin{align*} \label{eq:sfr}
\mathrm{log}\ \mathrm{SFR}\ (\mathrm{M_\odot\ yr}^{-1}) 
	&= \mathrm{log}\ \mathrm{\frac{L_{TIR}}{(L_\odot)}} - 9.83. \numberthis
\end{align*}

\noindent The uncertainty on the total infrared luminosities used for the SFR estimates is $\sim25\%$ \citep{Galametz:2013} and thus we suggest this as a lower limit to the uncertainty on the SFRs derived via Eq. \ref{eq:sfr} and listed in Table \ref{tab:ell_props}.

We estimate the dense molecular gas content, $\mathrm{M_{dense}}$, from the HCN luminosities using the conversion factor published by \cite{Gao:2004a}, $\mathrm{\alpha_{HCN}}\ \approx 10\ \mathrm{M_\odot\ (K\ km\ s^{-1}\ pc^{2})^{-1}}$. We discuss the possibility of variations in the HCN conversion factor in \S \ref{sec:conversion_factors}. To estimate the total molecular gas content, $\mathrm{M_{H_2}}$, we adopt a CO-to-$\mathrm{H_2}$ conversion factor of  $\mathrm{\alpha_{CO}}\ \approx 7\ \mathrm{M_\odot\ (K\ km\ s^{-1}\ pc^{2})^{-1}}$ from \cite{Schirm:2014}. \cite{Schirm:2014} estimate the CO abundance and conversion factor in the Antennae by modelling a warm and cold gas component using RADEX \citep{vanderTak:2007} and \textit{Herschel} Fourier Transform Spectrometer (FTS) data of multiple CO transitions. Using an initial CO abundance of $x_\mathrm{CO}\sim3\times10^{-4}$, they derive a warm H$_2$ gas mass that is $~10$ times lower than previous estimates from \cite{Brandl:2009} based on direct H$_2$ observations; assuming CO is tracing the same gas as H$_2$, they adjust their CO abundance to $x\sim5\times10^{-5}$. Using this abundance, their cold gas mass estimate is $\mathrm{M_{cold}}\sim1.5\times10^{10}$ M$_\odot$, resulting in the aforementioned CO $\mathrm{J}=1-0$ conversion factor. \cite{Wilson:2003} derive a similar conversion factor, $\mathrm{\alpha_{CO}}\ \approx 6.5\ \mathrm{M_\odot\ (K\ km\ s^{-1}\ pc^{2})^{-1}}$, by calculating the virial mass of resolved SGMCs using OVRO CO $\mathrm{J}=1-0$ data that we also use in this work.

Estimates of $\alpha_\mathrm{CO}$ from Milky Way observations show a factor $\sim5$ spread, with the typical value of $\mathrm{X_{CO}}\sim2\times10^{20}$ cm$^{-2}$ (cf. \citealt{Bolatto:2013}), which translates to $\mathrm{\alpha_{CO}}\sim4$ M$_\odot$ (K km s$^{-1}$ pc$^{2}$)$^{-1}$. \cite{Bolatto:2013} suggest a factor of $\sim2$ uncertainty for $\alpha_\mathrm{CO}$ applied to normal star-forming galaxies. Measurements of $\alpha_\mathrm{CO}$ in star-bursting galaxies show a spread of at least $\sim3$ (cf. \citealt{Bolatto:2013}), and are less studied and thus less well-constrained than $\alpha_\mathrm{CO}$ in more normal star-forming environments. Therefore, we suggest a factor $\sim4$ uncertainty on the mass estimates from $\alpha_\mathrm{CO}$. The HCN-to-dense H$_2$ conversion factor, $\alpha_\mathrm{HCN}$, is even less well-constrained than $\alpha_\mathrm{CO}$, so we suggest a factor of $\sim$10 uncertainty on the dense gas mass estimates.

We estimate the dense molecular gas fraction by taking the ratio of the dense molecular gas mass estimate from $\mathrm{L_{HCN}}$ to the total molecular mass mass from $\mathrm{L_{CO}}$, $\mathrm{f_{dense}} = \mathrm{M_{dense}/M_{H_2}}$. Similarly, we calculate the star formation efficiency of dense gas via $\mathrm{SFE_{dense} = SFR/M_{dense}}$ and the star formation efficiency of the total gas via $\mathrm{SFE = SFR/M_{H_2}}$. We calculate surface densities of the SFR, $\mathrm{M(H_2)_{dense}}$, and total $\mathrm{M(H_2)}$ by dividing these quantities by their elliptical aperture area (Table \ref{tab:ell_props}). The dense gas fraction should be uncertain by a factor of $\sim10$. The uncertainty on the star formation efficiencies is dominated by the mass uncertainties, and thus are also uncertain by a factor of $\sim10$ and $\sim4$ for SFE$_\mathrm{dense}$ and SFE, respectively.

\section{Discussion} \label{sec:discussion}

In this paper we study four distinct gas-rich star-forming regions in the Antennae: (1) the nucleus of NGC 4038, (2) the nucleus of NGC 4039, (3) the overlap region, and (4) the western arm. Our primary goals are to constrain the sub-galactic $\mathrm{L_{TIR}-L_{HCN}}$ relation in the Antennae, study how it varies across different regions, and piece together what drives this variation by characterizing the environment and star formation using line ratios and other results from the literature. {We include a list of luminosity ratios of regions in Table \ref{tab:ell_ratio} in Appendix \ref{app:lumin_ratio}.} In the following sections, we discuss the bulk properties of the Antennae observations and then discuss the $\sim$kpc-scale variations of different regions.


{\it A Note on SFRs from $\mathit{L_{TIR}}$: } In a merging system such as the Antennae, we can expect to be tracing a variety of stellar populations and SFRs. Models of mergers suggest there will be multiple bursts of star formation over the evolution of the system, with these bursts being triggered within different regions of the system at different times \citep{Mihos:1996, Mengel:2001}. Recent simulations estimate the Antennae system to be 40 Myr after its second pass \citep{Karl:2010}, placing it at a later stage in the Toomre sequence than previous estimates (e.g. \citealt{Toomre:1977}). This provides a natural explanation for the different ages and distributions of stellar populations observed across the Antennae, from bursts $<10$ Myr old to ancient globular clusters born in the progenitor galaxies ($\sim10$ Gyr, \citealt{Whitmore:1999}). Therefore, using SFR tracers at sub-galactic scales in this system requires caution, as the conversion from SFR-tracer luminosity to SFR may not be constant across the system.

The total infrared luminosity traces star formation over the past 100 Myr \citep{Kennicutt:2012} and so can be affected by the recent star formation history of a region. In particular, the $\mathrm{L_{TIR}}$ may underestimate the SFR in regions of young starbursts ($\le10$ Myr, \citealt{Brandl:2009, Kennicutt:2012}), or overestimate it in systems with a large population of evolved stars ($\ge100-200$ Myr) heating the dust. With a particularly violent episode of star formation ongoing in the overlap region, the first process could feasibly affect $\mathrm{L_{TIR}}$ SFR estimates in this region, while the second process may affect $\mathrm{L_{TIR}}$ SFR estimates in the nuclei and outer regions of the Antennae. However, these two effects would only act to \textit{enhance} the discrepancy we see in SFE$_\mathrm{dense}$ between the nuclei and overlap region of the Antennae. Furthermore, many regions in the Antennae galaxies are highly obscured by dust, and so other SFR tracers such as ultraviolet and H$\alpha$ are not reliable due to high extinction. $\mathrm{L_{TIR}}$ is commonly used as an SFR tracer in extragalactic studies and does not suffer from these extinction effects, making it a better SFR tracer in dusty environments. With these considerations,  we use $\mathrm{L_{TIR}}$ as our main SFR tracer, but we compare our results with other studies from the literature that target different stages of star formation using observations from other wavebands: radio continuum from the Very Large Array (VLA) \citep{Neff:2000}, X-ray from Chandra \citep{Zezas:2002}, optical/infrared from the Hubble Space Telescope (\textit{HST}) \citep{Whitmore:2010, Whitmore:1995}, and mm- and sub-mm observations from ALMA \citep{Whitmore:2014,Johnson:2015,Herrera:2017}. 

\subsection{General Characteristics of the Dense Gas: PDRs}

\cite{Leroy:2017} show that (for a fixed $\mathrm{T_{kin}}$) the total emissivity of a particular molecular transition is dependent on the width of the density PDF as well as the mean density at which it resides. This is such that even a molecule with a high critical density, like HCN, can emit brightly at low densities if the turbulence widens the density PDF sufficiently. They model emission from a number of molecular transitions, including HCN, HCO$^+$, HNC, and $^{12}$CO J$=1-0$, ratios using RADEX \citep{vanderTak:2007} for lognormal and lognormal+power law density distributions.  Throughout the majority of the Antennae, the integrated intensities of the dense gas lines we observe rank $\mathrm{HCO^+ > HCN > HNC}$. When \cite{Leroy:2017} vary only the mean density (and fix $\mathrm{T_{kin}}=25$ K), interestingly, the line ratios are ranked $\mathrm{HCO^+ > HNC > HCN}$ for low $\mathrm{n_0} < 10^3$ cm$^{-3}$, $\mathrm{HCN > HNC > HCO^+}$ for high $\mathrm{n_0} > 10^{3}$ cm$^{-3}$, or are all similar in strength at median densities $\mathrm{n_0} \sim 10^{3}$ cm$^{-3}$. This indicates that density variations alone cannot account for the difference in average lines strengths we observe, especially given that in all regions of the Antennae HNC emission is weaker than both HCO$^+$ and HCN. 

By default, RADEX takes into account the effect\footnote{These effects are included via source and sink terms in the statistical equilibrium calculations.} of chemical formation and destruction in the presence of cosmic-ray ionization plus cosmic-ray induced photodissociation on level populations. These rates are computed in a subroutine that can be modified to include a more complex treatment of chemical processes in molecular clouds. \citet{Loenen:2008} present models of Photon Dominated Regions (PDRs) and X-ray Dominated Regions (XDRs) that incorporate mechanical heating in addition to the PDR and XDR chemistry models of previous work (i.e. \citealt{Meijerink:2005,Meijerink:2007}). \citet{Loenen:2008} compare observed HCN, HCO$^+$, and HNC line ratios in nearby LIRGs to the results of the PDR and XDR modelling. Their results suggest that the HNC/HCN ratio is able to distinguish XDRs and PDRs, as this ratio never falls below unity for their X-ray dominated models. At higher temperatures ($>100$ K), HNC may be produced more efficiently than HCN in the HNC+H$\rightarrow$HCN+H reaction \citep{Schilke:1992,Talbi:1996}. Thus, in the presence of mechanical heating the HNC/HCN ratio is expected to be suppressed. As HNC is consistently weaker than HCN across the Antennae, the chemistry of the HNC- and HCN-emitting gas is likely UV-dominated (rather than X-ray dominated), with some amount of mechanical heating suppressing the HNC emission.

\citet{Loenen:2008} also suggest that the HCO$^+$/HCN and HCO$^+$/HNC ratios can distinguish high-density ($\mathrm{n>10^5}$ cm$^{-3}$) PDRs from lower-density PDRs. They divide the mechanical heating into: (1) stellar UV-radiation dominated chemistry arising from denser ($\mathrm{n>10^5}$ cm$^{-3}$) PDR environments with young ($<10$ Myr) star formation, resulting in HNC/HCN $\sim1$ and weak HCO$^+$, and (2) mechanical/supernovae-shock dominated chemistry from more diffuse PDR environments and stellar populations with ages $>10$ Myr, with HNC/HCN $<$ 1 and strong HCO$^+$ possible. \citet{Loenen:2008} attribute these differences in the HCO$^+$/HCN ratio primarily to differences in the density of the gas, rather than abundance variations. Since HCO$^+$ has a lower critical density than HCN and HNC, they argue that it is brighter than HCN and HNC in lower-density gas.  For the majority of the Antennae, HCO$^+$ is stronger than both HCN and HNC (except in NGC4038 and WArm-3 where HCN is actually stronger). Thus, the average ratios across the Antennae are consistent with the lower-density PDRs from \citet{Loenen:2008}, with some amount of shock heating from supernovae of $>10$ Myr stellar populations.

There is rough agreement between the lower-density models of \citet{Loenen:2008} and \cite{Leroy:2017} in the sense that HCO$^+$ is expected to be brighter than HCN. However, there are still large differences in the actual densities of the models that produce this trend; the lower-density \citet{Loenen:2008} models are at more moderate densities, $\mathrm{n}\sim10^{4.5}$ cm$^{-3}$, and the models of \cite{Leroy:2017} that produce this trend are $\mathrm{n}<10^{3.5}$ cm$^{-3}$. The models of \citet{Loenen:2008} (and \citealt{Meijerink:2005,Meijerink:2007}) assume a single density for the gas, although there is mounting evidence that variations in the gas density PDF can also significantly alter molecular luminosities (e.g \citealt{Leroy:2017}). Alternately, the \citet{Leroy:2017} models do not expand upon the default treatment of chemistry in the RADEX code. Therefore, future modeling of line-ratios should attempt to combine these treatments of gas density PDFs and chemistry for better constraints on the gas properties these line ratios are tracing in extreme environments.

\subsection{The nuclei of NGC 4038 and NGC 4039}

\subsubsection{Low $\mathit{L_{TIR}}/\mathit{L_{dense}}$ and High $\mathit{L_{HCN}}/\mathit{L_{CO}}$}
\label{sec:cmz_similarities}


The two nuclei appear to have lower $\mathrm{L_{TIR}}/\mathrm{L_{dense}}$ ratios compared to the ratios of the remaining Antennae regions (see Tables \ref{tab:ell_lumin}, \ref{tab:ell_props}, and \ref{tab:ell_ratio}, and Figure \ref{fig:ltir_lhcn_fit_all}). This ratio is taken as a proxy for the star formation efficiency of dense gas, SFE$_\mathrm{dense}$, assuming $\mathrm{L_{TIR}} \propto \mathrm{SFR}$ and $\mathrm{L_{dense}} \propto \mathrm{M_{dense}}$. The estimated SFE$_\mathrm{dense}$ for the nuclei are $0.44\times10^{-8}$ yr$^{-1}$ and $0.7\times10^{-8}$ yr$^{-1}$ for NGC 4038 and NGC 4039, respectively, compared to values of $1.1-4.6\times10^{-8}$ for regions in the overlap and western arm. We compare the $\mathrm{L_{TIR}}/\mathrm{L_{HCN}}$ ratios of the Antennae sources (from elliptical apertures) to those of \cite{Gao:2004a,Gao:2004b} and \cite{Liu:2015} in Figure \ref{fig:ltir_lhcn_fit_all}, which shows that the Antennae data, as a whole, span the majority of the range in $\mathrm{L_{TIR}}/\mathrm{L_{HCN}}$ ratios of these two samples of galaxies.  This also emphasizes the difference between the two nuclei (white stars) and the overlap and western arm regions (orange diamonds). The nuclei appear on the lower end of the locus of points from \cite{Gao:2004a,Gao:2004b} and \cite{Liu:2015}, below the median of the \cite{Gao:2004a,Gao:2004b} galaxies (850 $\mathrm{L_\odot}$ (K km s$^{-1}$ pc$^2$)$^{-1}$) while the overlap and western arm lie above this value in the typical starburst regime.

The two nuclei also show an enhancement in the $\mathrm{L_{HCN}/L_{CO}}$ ratio relative to the overlap region \citep{Schirm:2016}. This line ratio is often used as a proxy for f$_\mathrm{dense}$ ($\mathrm{L_{HCN}/L_{CO}\sim M_{dense}/M_{H_2}}$), which would indicate that the nuclei have higher dense gas fractions than other regions in the Antennae. Our calculated dense gas fractions are listed in Table \ref{tab:ell_props}, and indeed show that the nuclei have the highest dense gas fractions, with NGC 4038 at $\sim9.7\%$ and NGC 4039 at $\sim7.1\%$, compared to $2.2-4.5\%$ for sources in the overlap region. Regions in the western arm exhibit higher dense gas fractions up to $6.9\%$ in WArm-3, although these regions also show higher SFE$_\mathrm{dense}$ (e.g. SFE$_\mathrm{dense}=2.19\times10^{-8}$ yr$^{-1}$ in WArm-3), unlike the two nuclei.

This behavior in the nuclei (i.e lower SFE$_\mathrm{dense}$ and higher f$_\mathrm{dense}$) is similar to the inner regions of some disk galaxies (e.g. \citealt{Usero:2015, Bigiel:2016}), and could be attributed to an increase in ISM pressure. In some disk galaxies, f$_\mathrm{dense}$ has been observed to increase towards the center of the disk/bulge region (e.g. M51 in \citealt{Bigiel:2016} and the sample of galaxies in \citealt{Usero:2015}), and the star formation efficiency of the dense gas appears to decrease towards the center, showing an inverse correlation with f$_\mathrm{dense}$. This is also coincident with an increase in stellar density and gas fraction observed at the inner radii. If the gas is assumed to be in equilibrium with the hydrostatic pressure in the galaxy then increased ISM pressures arise naturally from this situation \citep{Helfer:1997, Hughes:2013, Bigiel:2016}.  Then, if the stellar potential were driving up the pressure in the nuclei, we might expect to see overall higher gas surface densities in these regions. Contrary to this, the nuclei appear to have moderate molecular gas surface densities ($\sim$240 and $\sim510$ M$_\odot$ pc$^{-2}$) compared to the SGMCs in overlap region ($\sim700-1000$ M$_\odot$ pc$^{-2}$). However, the surface density of the {\it dense} gas is higher in the NGC 4038 ($\sim50$ M$_\odot$ pc$^{-2}$) than in the overlap region ($\sim30$ M$_\odot$ pc$^{-2}$, while NGC 4039 shows $\Sigma_\mathrm{Mdense}=17$ M$_\odot$ pc$^{-2}$. We also note that dynamical equilibrium may not be a valid assumption in a merger system. For example, \cite{Renaud:2015} show that cloud-cloud collisions in the Antennae can increase pressure sufficiently through compressive turbulence to be able to produce massive cluster formation. Regardless of the source of increased pressure, it can potentially increase the dense gas content, as well as the mean density of the gas.  \cite{Loenen:2008} model line ratios in extreme environments such as (U)LIRGs (which are often merger systems) that are consistent with this.


Since the Antennae is a merger system, turbulent pressures are expected to be higher throughout this system (cf. \citealt{Renaud:2015}). Furthermore, the gas in any merger system will be drawn to the higher gravitational potential wells of the nuclei, thus potentially creating an even higher turbulent pressure in these regions. Turbulent pressure may also act to suppress star formation, and is the strongest candidate for explaining the star formation suppression in the Central Molecular Zone (CMZ, \citealt{Kruijssen:2014}). The CMZ is a region known to have high average gas densities ($\mathrm{n(H_2)>10^4}$ cm$^{-3}$, \citealt{Rathborne:2014}) despite a relative lack of star formation. \citet{Kruijssen:2014} have suggested that the lower SFR is attributed to an overall slower evolution of the gas towards gravitational collapse in the presence of higher turbulence (as the gas density threshold required for star formation is higher). Turbulence is the strongest candidate of the potential star formation suppressors in the CMZ (compared to tidal disruption, gas heating, etc.), and is likely due to gas inflow along the molecular bar or other disk instabilities \citep{Kruijssen:2014}. For this scenario to be true, the SFR in the CMZ must be episodic, suggesting that it is currently in a pre-starburst phase.  Other evidence suggests that clouds in the CMZ are not strongly self-gravitating \citep{Kauffmann:2017b}, but rather are being held together by the stellar potential. This also supports the idea that the SFR in the CMZ may increase in the future as gravitational collapse progresses in this region.

As mentioned previously, starburst episodes are natural in a merger system. If the nuclei are in a pre-(or post-) starburst phase, we may expect to measure higher ISM turbulent pressures from gas inflow (and/or stellar feedback). In the CMZ, pressures are $\mathrm{P/k_B}\sim10^9$ K cm$^{-3}$ \citep{Rathborne:2014}. Previous estimates of the pressure of the warm and cold components of lower-density gas in the Antennae (as measured by CO) show little variation and are $\mathrm{P/k_B}\sim10^5$ K cm$^{-2}$ \citep{Schirm:2014}. However, if mean gas densities are higher in the nuclei, then CO would not adequately trace the bulk properties of gas in these regions. Additionally, HCN/HCO$^+$ and HNC/HCN integrated line ratios differ between these sources, indicating that there may be different mechanisms driving the lower SFE$_\mathrm{dense}$ and f$_\mathrm{dense}$ in each of the two nuclei \citep{Schirm:2016} (NGC 4038 exhibits higher HNC/HCN and HCN/HCO$^+$ luminosity ratios than NGC 4039).  We investigate variation in the star formation between the two nuclei in \S \ref{sec:nuclei_sf}.

\citet{Rathborne:2014} argue that this lower SFR should be observed in the centers of other galaxies, and more evidence of this behavior is surfacing (e.g. \citealt{Usero:2015,Bigiel:2016}), with more work to come in the future. As discussed above, there are likely a number of sources of turbulence, including stellar feedback. In contrast, enhancements in the $\mathrm{SFE}$ of the total molecular gas content have been observed in the centers of some galaxies \citep{Utomo:2017}. \cite{Chown:2018} find enhancements in $\mathrm{SFE}$ and central gas concentrations in a number of barred and interacting galaxies, supporting the idea that mass transport can play a significant role in regulating star formation. However, the star formation history of these systems suggest that the enhancements have been sustained over long periods of time. More studies of the dense gas content in these systems will help determine if there is a common relationship between $\mathrm{SFE}$, $\mathrm{SFE_{dense}}$, and $\mathrm{f_{dense}}$ of the centers of barred and interacting galaxies. Overall, the parallels between the CMZ, the centers of disk galaxies, and the nuclei have interesting implications for star formation: processes affecting the SFR and gas PDFs of the CMZ and centers of disk galaxies may also be occurring in disturbed systems such as the Antennae. More work needs to be done to explore the mean density and density profile of the $\mathrm{L_{HCN}}$-emitting gas in these environments.

\subsubsection{Star Formation} \label{sec:nuclei_sf}

The two nuclei have the second and third highest L$_\mathrm{TIR}$ measurements in the system, below the L$_\mathrm{TIR}$ from SGMC345.  The SFRs we determine from our L$_\mathrm{TIR}$ measurements are 1.14 and 0.63 $\mathrm{M_\odot\ yr^{-1}}$, which are higher by a factor $>2$ than the estimates from \citet{Brandl:2009}.  \citet{Brandl:2009} use mid-IR fluxes (15 and 30 $\mu$m) to estimate L$_\mathrm{TIR}$, where we use the 24, 70, and 100 $\mu$m fluxes; to compare our estimates with theirs, we apply a scaling factor of 0.86 from \citet{Kennicutt:2012} to their SFR estimates based on the older \citet{Kennicutt:1998} SFR calibration\footnote{\citet{Kennicutt:2012} recommend multiplying L$_\mathrm{TIR}$-based SFR estimates using the \citet{Kennicutt:1998} calibration by a factor of 0.86.}. With the scaling factor applied, \citet{Brandl:2009} find SFRs to be $0.52$ and $0.27$ M$_\odot$ yr$^{-1}$ for NGC 4038 and NGC 4039, respectively, for L$_\mathrm{TIR}\sim3.67$ and $1.86\times10^9\ \mathrm{L_\odot}$.  Using the 70$\ \mu$m flux to estimate L$_\mathrm{TIR}$, \citet{Bigiel:2015} find L$_\mathrm{TIR}\sim8.8$ and $5.5\times10^9\ \mathrm{L_\odot}$ for their apertures Nuc. N and Nuc. S, although they do not convert these to SFRs. Our estimates for L$_\mathrm{TIR}$ are similar to this, with $\sim7.67$ and $4.22\times10^9\ \mathrm{L_\odot}$. It is likely our apertures are different than those used by \citet{Brandl:2009}, which may account for some of the differences. Regardless, NGC 4038 appears to have a SFR that is $\sim2$ times higher than NGC 4039. 

NGC 4039 has the characteristics of a post-starburst nucleus with little star formation activity. It hosts an older stellar population ($\sim65$ Myr from IR spectroscopic results/CO absorption, \citealt{Mengel:2001}) that is dominated by old giants and red supergiants (cf. photospheric absorptions line in the $\sim2\ \mu$m stellar continuum, \citealt{Gilbert:2000}).  \citet{Gilbert:2000} found no evidence of Br$\gamma$ emission, which is expected to be present in the atmospheres of young stars. NGC 4039 also has a steep radio spectrum \citep{Neff:2000} indicating that the radio emission is originating predominantly from SNe remnants of the $\sim65$ Myr-starburst. \textit{Chandra} observations reveal a composite X-ray spectrum that supports this picture: it contains a thermal component (indicating a hot ISM) and steep power-law with $\Gamma\sim2$ (indicating X-ray binaries, \citealt{Zezas:2002}). Furthermore, \citet{Brandl:2009} find evidence that H$_2$ in NGC 4039 is shock-heated. The lower HNC/HCN ratio we find in NGC 4039 is consistent with these findings and suggests it is driven by the mechanical heating of previous starbust activity and supernovae shocks \citep{Neff:2000}. 

\citet{Brandl:2009} find high excitation IR lines in the nucleus of NGC 4039, which is one potential indicator of an accreting stellar black hole binary. They measure a ratio [N\textsc{III}]/[N\textsc{II}]$\sim6$ times higher in NGC 4039 than NGC 4038 and strong [S \textsc{IV}], which was not detected in NGC 4038 at all. To determine the source of these high-excitation lines in NGC 4039, \citet{Brandl:2009} compare the mid-IR spectral continuum  ($\sim10-30\ \mu$m) to a starburst model template from \citet{Groves:2008}. They find it matches with a model representative of distributed star formation at solar metallicity, moderate pressures ($\mathrm{P/k_B\sim10^5\ K\ cm^{-3}}$), and a PDR fraction indicating star formation is still embedded. \citet{Brandl:2009} therefore interpret their line ratios as being consistent with dust emission heated solely by star formation.  The [S \textsc{IV}] emission in NGC 4039 may also trace young stars in a $\sim4-6$ Myr starburst. It is possible an episode of star formation may be in the very early stages in this nucleus. This is again consistent with the picture painted above for the CMZ, as the gas may be in a pre-starburst phase that will eventually go on to form stars at a higher rate. Br$\gamma$ emission is also detected in a circum-nuclear cluster (A1, \citealt{Gilbert:2007}) which is identified separately from and just north of NGC 4039; however, it falls within our aperture and is likely contributing to our SFR estimates in this region.

NGC 4038 also contains a post-starburst population aged at $\sim 65$ Myr \citep{Mengel:2001}. There is evidence of a younger $\sim6$ Myr population to the north of NGC 4038 \citep{Mengel:2001}. NGC 4038 has a very soft X-ray spectrum likely due to thermal emission originating from winds from this region of young star formation \citep{Zezas:2002}. Br$\gamma$ emission is detected in the northern nucleus, which provides evidence for young star formation in this region. The X-ray luminosity of NGC 4038 is also lower than that of NGC 4039. Thus, the star formation in NGC 4038 is at a different stage than NGC 4039, and may be at the upswing of a starburst.

\subsection{The Western Arm}

\citet{Whitmore:2010} show that the Antennae presents an interesting number of large- and small-scale patterns related to star formation. One of these regions with such patterns is the western arm. \citet{Whitmore:2010} study the population of star clusters in the Antennae using \textit{Hubble Space Telescope} images from ACS and NICMOS. In the western arm, they designate five knots of clusters (originally discovered by \citealt{Rubin:1970}) that spatially coincide with dense gas emission detected in our study; sources G, L, R, S, and T overlap with our apertures WArm-1 (G), WArm-3 (T, S, and R), and WArm-4 (L). In their study, they note linear spatial age gradients in several clusters, including knots S, T,  and L in the western arm. The ages appear to increase towards the inner side of the spiral pattern in the direction of major dust lanes. The dense gas detected along the western arm also appears to be concentrated on the inner portion of the spiral pattern coincident with the dust lanes in this region, excluding WArm-1. (WArm-1 appears more centralized in the northern portion of the spiral pattern.) \citet{Whitmore:2010} posit that this gradient may be due either to small-scale processes, such as sequential star formation, or larger-scale processes such as density waves or gas cloud collisions. Either of these processes could also explain the position of the dense gas emission towards the inner portion of the western arm. 

The western arm hosts several bright H\textsc{II} regions, as evidenced in the H$\alpha$ image from HST (Fig. \ref{fig:data}). The diameters of these hot bubbles are widest along the western arm, indicating slightly more evolved starbursts than the overlap region \citep{Whitmore:2010}. The HCN/HCO$^+$ ratio varies from $\sim0.7-1.3$ in this region, with the highest ratio exceeding unity in WArm-3. 

{\it WArm-3: } The dense gas emission associated with WArm-3 overlaps with the inner edge of the H\textsc{II} region associated with knot S, and likely originates from gas shock-heated by UV winds and SNe. This region also shows bright compact 4- and 6-cm emission with both shallow and steep spectral indices \citep{Neff:2000}, indicating a combination of thermal emission and synchrotron emission from SN remnants, which could potentially be from the exposed O-star remnants. This is one of the few regions in the Antennae where HCN emission exceeds HCO$^+$ (HNC remains very weak/undetected), with HCN also appearing more spatially-extended. The abundance of HCO$^+$ can be significantly reduced in environments with a high ionization fraction, while the HCN abundance remains relatively unaffected \citep{Papadopoulos:2007}. This could potentially account for the higher HCN/HCO$^+$ ratio here, considering the proximity of this dense gas emission to the H\textsc{II} regions of knots R, S, and T. Or, this region could simply be at an overall higher-density, with mechanical heating continuing to drive down the HNC abundance. \citet{Brandl:2009} also study this starbursting region in the western arm (their Peak 4). Age estimates place the stellar population here around $\sim7$ Myr \citep{Whitmore:2002,Mengel:2005,Brandl:2009}.  Again, their SFR (0.22 M$_\odot$ yr$^{-1}$) agrees with ours to within $50\%$. 

{\it WArm-2: } The WArm-2 region, like WArm-3, is coincident with an obscuring dust lane (see Fig. 1). This region does not appear to have compact cm-emission or optical knots associated with it \citep{Neff:2000, Whitmore:2010}, and it also appears to have the lowest estimated SFR in our sample (0.1 $\mathrm{M_\odot}$ yr$^{-1}$). There are a few compact H\textsc{II} regions associated with this region (visible in H$\alpha$, \citealt{Whitmore:2010}), indicative of younger, embedded star formation. There is HCO$^+$ emission associated with this region, but no detected HCN or HNC. This is consistent with a lower mean density of gas, as both species have higher critical and effective densities than HCO$^+$ \citep{Shirley:2015}. It is interesting that HCN is not detected despite there being evidence of star formation in this region. This may indicate that HCN is not an efficient tracer of dense gas at this particular stage of star formation, or perhaps other mechanisms are suppressing the HCN emission that currently remain unclear. It is possible that one or more of these transitions are optically thick and subject to radiative trapping. For example, \cite{JimenezDonaire:2017} show that radiative trapping can effectively reduce the critical density required to stimulate $^{12}$C-transitions, thus boosting the intensity of these lines relative to $^{13}$C-transitions. Something similar could potentially occur between HCO$^+$, HCN, and HNC where one or more of these transitions is boosted relative to the other from optical depth variations. However, for optical depth variations to explain HCO$^+$ being detected over HCN, HCO$^+$ would need a higher optical depth and higher critical density than HCN, which we find unlikely.

Another possible explanation for the detection of HCO$^+$ over HCN and HNC is that nitrogen is possibly depleted in WArm-2. A mechanism for this would be low-metallicity gas flowing into the western arm from the outskirts of the galaxy. However, this should affect all regions in the western arm equally, and HCN is detected in WArm-1 and WArm-3. In fact, HCN is brighter than HCO$^+$ in WArm-3. Therefore, we find it more likely that the variations we observe are due to excitation effects, such as density variations.

{\it WArm-1: } North of WArm-2 is WArm-1, which has visible HCN and HCO$^+$ emission. The line ratios for this source are consistent with the average line ratios of the entire system: HCO$^+$ is brighter than HCN, and HNC is relatively weak/not detected. There also appear to be optical clusters associated with this region, in particular knot G from \cite{Whitmore:2010}. \citet{Neff:2000} detect compact radio emission in the vicinity of WArm-1 (their region 13), of which five sources have detections at both 4 and 6 cm and allow for the estimation of their radio spectral indices. Three of these sources have indices $>-0.4$, indicating strong thermal sources, while two have steep non-thermal emission indicated by indices $\sim -0.45$ and $\sim -1.64$.  \citet{Zezas:2002} detect 18 ultra-luminous X-ray (ULX) sources ($\mathrm{L_X}>10^{39}$ erg s$^{-1}$) in the Antennae, which they suggest are accreting black hole binaries. One of these sources, their X-ray source 16, is also coincident with the dense gas emission in WArm-1. This is one of three variable ULX sources, which further supports the idea that these are black hole binaries.

{\it WArm-4: } The WArm-4 region appears at the southern tail of the spiral pattern. The ratios in this region also follow the average trend of the system. Again, optical clusters (knot L, \citealt{Whitmore:2010}) and compact radio emission (region 8, \citealt{Neff:2000}) are associated with this region. The region of compact radio emission associated with knot L has a spectral index that indicates thermal emission is the dominant source ($\sim 0.18$, \citealt{Neff:2000}). There appear to be no medium/hard X-ray sources associated with this region, although there is diffuse soft X-ray emission throughout the Antennae \citep{Zezas:2002}.

\subsection{The Overlap Region \label{sec:overlap_SF}}


We find our SFR estimates are systematically lower for the SGMCs in the overlap region than the estimates from \citet{Brandl:2009}, perhaps related to the different TIR calibrators used. For clouds in the overlap region, \citet{Brandl:2009} estimate 0.61 M$_\odot$ yr$^{-1}$ for SGMC 1 (their Peak 3, corrected) and 3.14 M$_\odot$ yr$^{-1}$ for SGMC345 (the addition of their measurements for their Peaks 1 and 2, corrected). Our estimate for SGMC 2 (their Peak 5), 0.63 M$_\odot$ yr$^{-1}$, agrees well with their value of 0.55 M$_\odot$ yr$^{-1}$ (corrected).  \citet{Brandl:2009} suggest that L$_\mathrm{TIR}$ estimates may be high for regions with stellar populations $<10$ Myr; this in particular would affect measurements in the overlap region, which contains stellar populations as young as $\sim2-5$ Myr \citep{Brandl:2009, Whitmore:2002, Mengel:2005, Gilbert:2007, Snijders:2007}. Additionally, mid-IR fluxes are more sensitive to younger stellar populations \citep{Kennicutt:2012}, and these wavelengths may be better tracers of star formation in the overlap region; this could explain the discrepancy between our measurements (from 24, 70, and 100 $\mu$m IR observations) and those from \citet{Brandl:2009}, implying our estimates may be low.

There are numerous studies on star formation in the overlap region, with recent high-resolution ALMA studies now revealing the formation of super-star clusters (SSCs) \citep{Whitmore:2010,Johnson:2015,Herrera:2017}. At higher resolution, it is easier to distinguish the individual SGMCs, their associated clusters, and the conditions accompanying them. Theoretical studies of SSCs suggest that they require high pressures ($\mathrm{P/k_B}\sim10^7 - 10^8$ K cm$^{-3}$) are required for their formation \citep{Herrera:2017}. As mentioned above, \cite{Schirm:2014} found moderate pressures across the entire system, $\mathrm{P/k_B}\sim10^5$ K cm$^{-3}$ using an excitation analysis of multiple-$J$ transitions of CO.  However, these observations were using lower-resolution data ($\sim43''$) which likely will not capture the conditions necessary to form SSCs, since these form on much smaller scales. The existence of SSCs in the overlap region strongly suggests that gas pressures are  higher than these previous estimates.  

Our apertures in the overlap region coincide with bright star-forming knots B (our aperture  SGMC345), C, and D (\citealt{Rubin:1970, Whitmore:2010}, our SGMC1), and a more extended star-forming region 2 (\citealt{Whitmore:1995}, our SGMC2, C6, C7, and C8*). Our C9* region is adjacent (west) to star-forming knot B and does not coincide with bright optical star-forming regions. The strongest thermal radio source in \citet{Neff:2000} lies in the overlap region and falls within our aperture SGMC345. More specifically, this thermal source is overlapping with SGMCs 4 and 5, with SGMC 3 off further to the west. \citet{Neff:2000} estimate that $\sim5000$ O5 stars would be required to ionize this gas, resulting in an absolute magnitude of $-15$, or 500,000 B0 resulting in a magnitude of $-18$, bright enough to be detected with \textit{HST} if the starlight is not obscured by foreground dust or gas. However, \cite{Whitmore:1995} do not detect bright cluster emission near this radio source. Therefore, \citet{Neff:2000} suggest that star formation must be embedded in this particular complex, hidden by optical extinction that is at least 4 orders of magnitude. We measure the highest SFR in SGMC345, 1.46 $\mathrm{M_\odot}$ yr$^{-1}$, which is consistent with this  being the most vigorously star-forming complex in the Antennae.

\subsection{Conversion Factors} \label{sec:conversion_factors}


We use the $\mathrm{L_{HCN}/L_{CO}}$ ratio as an estimator of dense gas fraction across the Antennae assuming constant conversion factors, $\alpha_\mathrm{HCN}$ and $\alpha_\mathrm{CO}$. However, if $\alpha_\mathrm{HCN}$ and $\alpha_\mathrm{CO}$ vary across the Antennae, the trends we see between $\mathrm{L_{TIR}}$, $\mathrm{L_{HCN}}$, and $\mathrm{L_{CO}}$ may not be a consequence of different dense gas fractions. In particular, the CO conversion factor can vary with several gas properties, including metallicity, CO abundance, temperature, and gas density variations (cf. \citealt{Bolatto:2013}). Using computational models, \cite{Narayanan:2011} study the effects of varying physical properties on $\alpha_\mathrm{CO}$ in disks and merging systems, and they find $\alpha_\mathrm{CO}$ is typically lower in regions of active star formation in merger-driven starbursts. This is primarily due to higher gas temperatures and larger gas velocity dispersions in these systems (from increased thermal dust-gas coupling). They also show that $\alpha_\mathrm{CO}$ can either stay low or rebound after the starburst phase ends, depending on H$_2$ or CO abundances and the time required to revirialize gas. If we extrapolate these results to the $\sim$kpc scales studied in the Antennae, one would expect the overlap region to have a smaller $\alpha_\mathrm{CO}$ than the two nuclei, as this is the most vigorously star forming region in the merger. 

However, \cite{Zhu:2003} find evidence that the CO conversion factor may be 2-3 times lower in NGC 4038 than in the overlap region using Large Velocity Gradient (LVG) modelling of multiple $^{12}$CO and $^{13}$CO transitions. They find $\mathrm{X_{CO}}\sim(5.1-6.4)\times10^{19}(10^{-4}/{x_\mathrm{CO}})$ cm$^{-2}$ (K km s$^{-1}$)$^{-1}$ in the overlap region and $\mathrm{X_{CO}}\sim2.3\times10^{19}(10^{-4}/{x_\mathrm{CO}})$ cm$^{-2}$ (K km s$^{-1})^{-1}$ for NGC 4038, where $\mathrm{X_{CO}}$ is the two-dimensional conversion factor, $\Delta\mathrm{V}$ is the line width, and $x_\mathrm{CO}$ is the CO abundance relative to H$_2$ (the CO abundance is typically $x_\mathrm{CO}\sim10^{-5}-10^{-4}$ in starbursts, \citealt{Booth:1998,Mao:2000}). \cite{Zhu:2003} argue that the lower conversion factor in NGC 4038 is due to high velocity dispersion, large filling fraction, and low optical depth of the CO-emitting gas. \citet{Sandstrom:2013} show that the CO conversion factor is lower by a factor of $\sim2$ (on average) in the central 1 kpc of a sample of 26 star-forming disk galaxies, and they also find that it can be up to 10 times lower than the standard Milky Way value ($\alpha_\mathrm{CO} = 4.4$ M$_\odot$ pc$^{-2}$ (K km s$^{-1}$)$^{-1}$)). \citet{Sandstrom:2013} suggest several explanations for this discrepancy in $\alpha_\mathrm{CO}$ in the central regions of these galaxies, including differences in ISM pressure, higher molecular gas temperatures, and/or more diffuse ISM molecular gas; optical depth variations can also alter the conversion factor of the gas and can act in accordance with any of the previous effects.

It is possible that the effects on $\alpha_\mathrm{CO}$ in  galaxy centers observed in the \citet{Sandstrom:2013} sample could still apply to the two nuclei in the Antennae. If $\alpha_\mathrm{CO}$ is indeed lower in NGC 4038, this would \textit{decrease} the total molecular mass estimates in that nucleus and would \textit{increase} the dense gas fraction in NGC 4038, thus exacerbating the difference between this nucleus and the overlap region. Metallicities have been found for young and intermediate stellar clusters across the Antennae ranging from slightly sub-solar to super-solar (Z$=0.9-1.3$ Z$_\odot$, \citealt{Bastian:2009}), but there are no obvious differences between clusters near the two nuclei vs those in the overlap region. High-resolution observations resolving gas at $\sim$kpc scales of multiple CO transitions have yet to be done in the Antennae and therefore gas properties such as density, temperature, and abundance are not constrained at these scales.


The conversion between HCN luminosity and dense gas mass is not as well studied as the CO conversion factor, but the HCN conversion factor is derived using the same principles assumed for $\alpha_\mathrm{CO}$. Thus, $\alpha_\mathrm{HCN}$ may also change with metallicity, pressure, temperature, density, etc. Observations of HCN and HCO$^+$ lines in (U)LIRGs suggest that HCN can experience a large range of excitation conditions (e.g. \citealt{Papadopoulos:2007,Papadopoulos:2014}), with some of these extreme galaxies showing sub-thermal HCN emission (i.e. Mrk 231, \citealt{Papadopoulos:2007}). Galactic observations of HCN in Orion A also show that HCN can be excited at more moderate densities, $\mathrm{n}\sim10^3$ cm$^{-3}$ \citep{Kauffmann:2017c}, missing dense gas entirely in some star-forming environments. \cite{Shimajiri:2017} directly compare HCN $\mathrm{J}=1-0$ emission and dust column density maps in three Galactic star-forming clumps and find evidence that $\alpha_\mathrm{HCN}$ may decrease with increasing local FUV radiation field, $\mathrm{G}_0$. In an attempt to calibrate $\alpha_\mathrm{HCN}$ numerically, \cite{Onus:2018} study the dependence of $\alpha_\mathrm{HCN}$ on different physical conditions using simulations of star-forming gas at $\sim2$ pc scales. They find that variations in HCN abundance ($3.3\times10^{-9}$ vs $3\times10^{-8}$) change $\alpha_\mathrm{HCN}$ by a factor of $\sim2$, and that moderate differences in temperature (10 vs 20 K) can also alter $\alpha_\mathrm{HCN}$, but less significantly. So far these observations and simulations have been limited to small spatial scales (on the order of $\sim$10 pc). At these smaller scales, there is some expected stochasticity of physical conditions within molecular clouds that may affect $\alpha_\mathrm{HCN}$. For applications to extragalactic observations, this work needs to be expanded to larger scales (kiloparsecs) to better estimate $\alpha_\mathrm{HCN}$. 

\subsection{Dense Gas Fractions} \label{sec:fdense}

Traditionally, the dense gas fraction is estimated as the direct ratio of total molecular mass (traced by CO) to dense gas mass (traced by HCN). This works under the assumption that CO is tracing the mean density of gas, while HCN is tracing only the high-density gas that is more directly associated with star formation. However, recent work on the CMZ suggests that this may be an oversimplification; \cite{Kruijssen:2014} show that the overall gas density PDF can be pushed to significantly higher densities via turbulence, while the gas densities more directly associated with star formation are even higher \citep{Rathborne:2014}. In this regime, HCN is a better tracer of the mean density of gas. Therefore, our interpretation of the $\mathrm{L_{TIR}}/\mathrm{L_{dense}}$ and $\mathrm{L_{HCN}}/\mathrm{L_{CO}}$ ratios may vary depending on the regime of star formation we are in. 

Current turbulent models of star formation predict lognormal gas density PDFs that evolve to have power-law tails once gravitational collapse begins in the process of star formation (e.g. \citealt{Federrath:2012}). Gravitational collapse will begin once the gas reaches a threshold density that is high enough to overcome pressure support in the cloud. If the source of pressure is turbulence, it can act to (1) widen the gas density PDF, and/or (2) push the overall mean density of the gas to higher values \citep{Federrath:2012}. In this case, we may expect to see an enhancement of luminosities of dense gas tracers, such as HCN, relative to lower-density gas tracers such CO (cf. \citealt{Leroy:2017}).

We suggested in \S \ref{sec:cmz_similarities} that similar effects of the gas and star formation in the CMZ may be affecting the two nuclei in the Antennae. If the gas density PDF is indeed shifted to higher densities in the nuclei, this may result in a \textit{smaller} HCN conversion in these regions, which would \textit{decrease} the dense gas fraction in these regions. For example, turbulence in the CMZ has the effect of driving up the mean density of gas to $\mathrm{n(H_2)\sim}10^4$ cm$^{-3}$, which is 100 times larger than the mean density of GMCs elsewhere in the Milky Way \citep{Rathborne:2014}. Similarly, the threshold of gas required for star formation (also referred to as a critical density in some literature) in the CMZ is also higher, $\mathrm{n(H_2)_{thresh}\sim}10^6$ cm$^{-3}$ \citep{Rathborne:2014}. Therefore, an accurate measure of dense gas fraction in this region is a comparison of the mass at densities $>10^6$ cm$^{-3}$, M($>10^6$ cm$^{-3}$), to that of the total molecular gas content. In the CMZ, HCN can be well-excited already at $\mathrm{n(H_2)\sim}10^4$ cm$^{-3}$, the mean density of the gas. This makes HCN a better tracer of the \textit{mean} density of gas in the CMZ, rather than CO. Similar effects likely affect the luminosity measurements in the nuclei of the Antennae. 

\section{Conclusion} \label{sec:conclusion}

We present a study of the dense gas content and star formation in NGC 4038/9, with detections of HCN, HCO$^+$, and HNC J=1-0 emission in four distinct regions of the Antennae: the two nuclei (NGC 4038, NGC 4039), the overlap region, and the western arm. We consider the two nuclei separately as they exhibit differences in dense gas line ratios and star formation activity. 

\begin{enumerate}
	\item The two nuclei show a suppression in the $\mathrm{L_{TIR}/L_{HCN}}$ ratio, despite showing an enhanced $\mathrm{L_{HCN}/L_{CO}}$ ratio, when compared with the overlap and western arm regions. Assuming constant conversion factors, $\alpha_\mathrm{HCN}$ and $\alpha_\mathrm{CO}$, this suggests the two nuclei have a higher dense gas fraction and lower star formation efficiency of dense gas compared to the rest of the Antennae. One potential explanation for this is an increase in overall turbulence in these regions that acts to suppress star formation while also increasing the overall gas density, similar to what appears to be happening in the CMZ in the Milky Way \citep{Kruijssen:2014}. This behavior is expected in the pre-starburst phase of merger systems \citep{Narayanan:2011}.
	\item The Antennae data extend the $\mathrm{L_{TIR}\ vs.\ L_{HCN}}$ relationship observed by \citet{Gao:2004a,Gao:2004b} to lower luminosity, consistent with the results from \citet{Bigiel:2015}.   The Antennae datapoints fit within the scatter of the \citep{Gao:2004a,Gao:2004b} and \cite{Liu:2015} datapoints. A fit of the Antennae data with that of \citeauthor{Liu:2015} results in a power-law index of m$\sim$1.
	\item A fit to the Antennae $\mathrm{L_{TIR}}$ and $\mathrm{L_{HCN}}$ data shows a sub-linear relationship with a power law index m$\sim$0.5 (hexagonal apertures). Fits with $\mathrm{L_{HCO^+}}$ and $\mathrm{L_{HNC}}$ similarly show sub-linear power law indices of m$\sim$0.5 and m$\sim$0.6, respectively. Assuming L$_\mathrm{TIR}\sim$SFR and $\mathrm{L_{HCN}}\sim$M$_\mathrm{dense}$, this indicates of variations in the star formation efficiency of dense gas across this system, such that SFE$_\mathrm{dense}$ does not increase directly with M$_\mathrm{dense}$.  
	\item Except for NGC 4038 and WArm-3, the HCN/HCO$^+$ ratio is less than unity for regions in the Antennae, and HNC is significantly weaker than HCN and HCO$^+$. These average line ratios of HCN, HCO$^+$, and HNC are consistent with a lower-density ($\mathrm{n<10^5}$ cm$^{-3}$) PDR dominated by mechanical heating from stellar UV- and SNe shock-driven chemistry \citep{Loenen:2008}. 
	\item \cite{Schirm:2016} revealed bright, dense gas emission in the overlap region and two nuclei. At the tapered resolution of this study, bright HCN and HCO$^+$ J$=1-0$ emission is also detected along the inner portion of the western arm of the Antennae, which also coincides with a dust lane. Stellar clusters show age gradients, increasing in age towards the inner portion of the arm. Since this coincides with bright dense gas emission and dust, this supports the idea that the clusters formed via sequential star formation, or larger-scale processes such as density waves or gas cloud collisions \citep{Whitmore:2014}.
\end{enumerate}

\section*{Acknowledgements}

We thank the referee for their comments which have helped to improve the paper. This paper makes use of the following ALMA data: ADS/JAO.ALMA$\#$2012.1.1.00185.S. ALMA is a partnership of ESO (representing its member states), NSF (USA) and NINS (Japan), together with NRC (Canada), MOST and ASIAA (Taiwan), and KASI (Republic of Korea), in cooperation with the Republic of Chile. The Joint ALMA Observatory is operated by ESO, AUI/NRAO and NAOJ. The National Radio Astronomy Observatory is a facility of the National Science Foundation operated under cooperative agreement by Associated Universities, Inc. This work is based in part on observations made with the Spitzer Space Telescope, which is operated by the Jet Propulsion Laboratory, California Institute of Technology under a contract with NASA. Part of this work is based on observations made with \textit{Herschel}. Herschel is an ESA space observatory with science instruments provided by European-led Principal Investigator consortia and with important participation from NASA. AB wishes to acknowledge partial support from an Ontario Trillium Scholarship (OTS). CDW acknowledges financial support from the Canada Council for the Arts through a Killam Research Fellowship. The research of CDW is supported by grants from the Natural Sciences and Engineering Research Council of Canada and the Canada Research Chairs program. This research made use of Astropy, a community-developed core Python package for Astronomy \cite{astropy}. A significant amount of this research also made use of Matplotlib \citep{matplotlib}, Photutils \citep{photutils}, Numpy \citep{numpy}, and Pandas \citep{pandas} Python packages.  This research has made use of the NASA/IPAC Extragalactic Database (NED) which is operated by the Jet Propulsion Laboratory, California Institute of Technology, under contract with the National Aeronautics and Space Administration. This research has made use of NASA's Astrophysics Data System Bibliographic Services.


\appendix
\renewcommand{\thesubsection}{\thesection.\Roman{subsection}} 


\section{Uncertainties} \label{app:uncertainties}

\subsection{Molecular Luminosities} \label{app:mol_lumin}

There are three primary sources of uncertainty on the molecular luminosities that we consider: 1. the calibration uncertainty of the ALMA data ($\sim 5\%$ for Band 3, ALMA Technical Handbook for Cycle 1), 2. the rms uncertainty of the moment zero maps, and 3. the uncertainty on the luminosity distance from \cite{Schweizer:2008} that we use. The rms uncertainty per aperture is discussed in \S \ref{sec:analysis} and given by equation \ref{eq:ap_noise}. There is a $5\%$ flux calibration uncertainty on each pixel that adds with the distance and aperture rms uncertainties in quadrature:

\begin{align*}
\sigma_\mathrm{L'} &= \mathrm{L'} \sqrt{ \left( \mathrm{\frac{\sigma_{M_{0},ap}}{M_{0,{ap}}}} \right)^2 + \mathrm{(0.05)^2} + 2\left( \mathrm{\frac{\sigma_{D_L}}{D_L}} \right)^2 }
\end{align*}


\subsection{Infrared Measurements and Luminosities} \label{app:ltir_uncer}

Uncertainty maps of the \textit{Herschel} and \textit{Spitzer} data are included in each of the downloaded fits files (created by the relevant reduction software) and describe the instrumental uncertainty, such that each pixel has an associated value and uncertainty: $\mathrm{S}_\nu\mathrm{(x,y}) \pm \sigma_{\mathrm{inst},\nu}\mathrm{(x,y)}$. Each of the instruments have a flux calibration uncertainty that also needs to be folded into the total uncertainty estimate of each pixel, $\sigma_\mathrm{cal}$, which are 5\%, 5\%, and 4\% for PACS \citep{Poglitsch:2010}, SPIRE \citep{Griffin:2010}, and MIPS \citep{Bendo:2012}, respectively. For each of the IR maps, we estimate the background level using three, separate apertures selected within the flat region of the background in each map. The average of this background level is subtracted from our measurements, and the corresponding background-subtraction uncertainty, $\sigma_\mathrm{back,\nu}$, is also folded into our final measurement uncertainties. The absolute uncertainty on the flux in a single pixel, $\mathrm{S}_\nu$, can be written:
\begin{align*}
\sigma_{\mathrm{S}_\nu}\mathrm{(x,y)} &= \sqrt{\sigma_{\mathrm{inst},\nu}\mathrm{(x,y)}^2 + (\sigma_\mathrm{cal}\times\mathrm{S}_\nu\mathrm{(x,y)})^2 +\sigma_\mathrm{back,\nu}^2}
\end{align*}

\noindent We convert IR fluxes to single-band luminosities using Eq. \ref{eq:lir_watts}, so the absolute uncertainty on these luminosities at pixel (x,y) can be written (incorporating the distance uncertainty) as:

\begin{align*}
\mathrm{ \sigma_{\nu L_\nu}(\mathrm{x,y}) = \nu L_\nu(\mathrm{x,y}) \sqrt{2\left(\frac{\sigma_{d_L}}{d_L}\right)^2 + \left(\frac{\sigma_{S_\nu}(\mathrm{x,y})}{S_\nu(\mathrm{x,y})}\right)^2} }
\end{align*}

The monochromatic $\mathrm{L_{TIR}}$ calibrations from \citet{Galametz:2013} are given in the form $\mathrm{log(L_{TIR}) = {a_i {log}(\nu_i L_{\nu,i}) + b_i}}$, where $\mathrm{ a_i\pm\sigma_{a_i} }$ and $\mathrm{ b_i\pm\sigma_{b_i} }$ are fit parameters and their uncertainties for IR band i. To derive the uncertainty on $\mathrm{log(L_{TIR})}$, we use standard error propagation:

\begin{align*}
\sigma_{\mathrm{log(L_{TIR})}} &=  \mathrm{ \sqrt{ \left( log(\nu L_{\nu,i}) \sigma_{a,i} \right)^2 +  \sigma_{b,i}^2 + \left( \frac{a_i \sigma_{L_{\nu,i}}}{ L_{\nu,i} ln(10)} \right)^2 }}
\end{align*}

\noindent where the absolute uncertainty on the total infrared luminosity is then: 

\begin{align*}
\sigma_{L_\mathrm{TIR}} = L_\mathrm{TIR} \mathrm{ln}10 \sigma_{\mathrm{log} L_\mathrm{TIR}}
\end{align*}

\noindent The \cite{Galametz:2013} calibrations combining more than one IR band are in the form $\mathrm{L}_\mathrm{TIR} = \sum_{i}{c_i \nu L_\nu(i)}$ where $c_i\pm\sigma_{c_i}$ are the fit parameters and their uncertainties for band $i$. The uncertainties on $\mathrm{L}_\mathrm{TIR}$ are then:

\begin{align*}
\sigma_{\mathrm{L}_\mathrm{TIR}} &= \sqrt{\sum_{i}\left(c_i \sigma_{\nu L_\nu(i)} \right)^2 + \left(\nu L_\nu(i) \sigma_{c_i} \right)^2}
\end{align*}

\noindent \citet{Galametz:2013} suggest an uncertainty of $\sim50\%$ on the monochromatic $\mathrm{L}_\mathrm{TIR}$ estimates, and an uncertainty of $\sim30\%$ on those combining multiple bands. Therefore, we cite the uncertainties that are largest (i.e. the percentage uncertainty from \citet{Galametz:2013} vs. our absolute uncertainty derivations).



\section{$\mathrm{L_{TIR}}$ Calibrations} \label{app:ltir}

\begin{figure*}[tb]
    \centering
    \includegraphics[width=\textwidth]{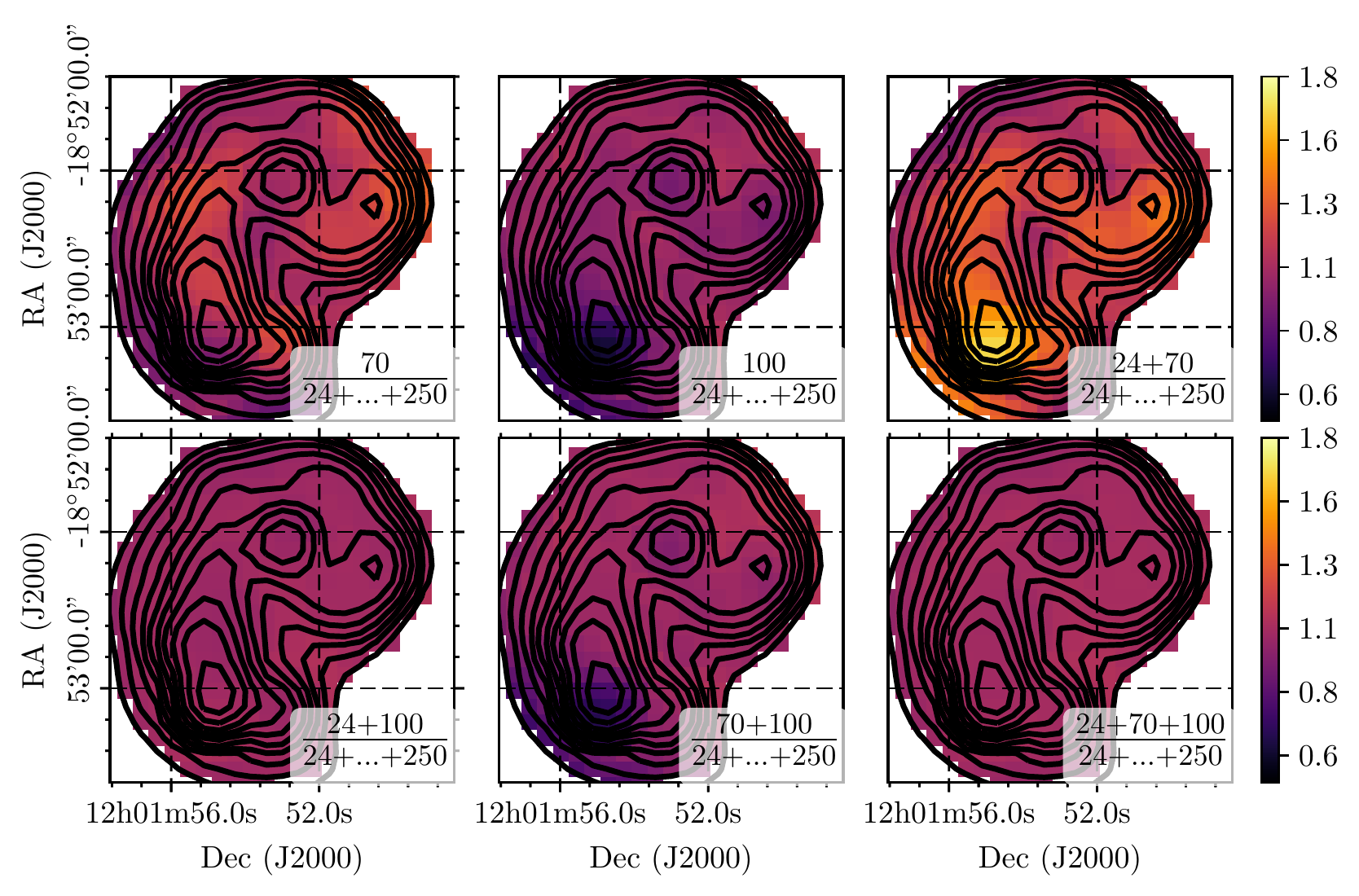}
    \caption{Ratio maps of \cite{Galametz:2013} calibrations at the 250 $\mu$m resolution ($18.1''$) with 250 $\mu$m contours overlaid. The $\mathrm{L_{TIR}}(24+70+100+160+250)$ calibration puts the tightest constraints on the total infrared luminosity estimate, therefore we compare combinations of calibrations with just the higher-resolution IR data (i.e. 24, 70, and 100 $\mu$m maps) to this calibration. \textit{Top, left to right:} $\mathrm{L_{TIR}}(70)/\mathrm{L_{TIR}}(24+70+100+160+250)$, $\mathrm{L_{TIR}}(100)/\mathrm{L_{TIR}}(24+70+100+160+250)$, $\mathrm{L_{TIR}}(24+70)/\mathrm{L_{TIR}}(24+70+100+160+250)$.  \textit{Bottom, left to right:} $\mathrm{L_{TIR}}(70+100)/\mathrm{L_{TIR}}(24+70+100+160+250)$, $\mathrm{L_{TIR}}(24+100)/\mathrm{L_{TIR}}(24+70+100+160+250)$, $\mathrm{L_{TIR}}(24+70+100)/\mathrm{L_{TIR}}(24+70+100+160+250)$. }
    \label{fig:ltir250}
\end{figure*}

We expect the $\mathrm{L_{TIR}}(24+70+100+160+250)$ calibration from \citet{Galametz:2013} puts the tightest constraints on the total infrared luminosity estimate since it most precisely reproduces the modelled $\mathrm{L_{TIR}}$ estimates in \citet{Galametz:2013} in comparison to the calibrations using fewer bands. Therefore, we compare other calibrations with just the higher-resolution IR data (i.e. 24, 70, and 100 $\mu$m maps) to this calibration to assess their spatial variation across the Antennae. We show ratios of these $\mathrm{L_{TIR}}$ calibrations to the $\mathrm{L_{TIR}}(24+70+100+160+250)$  at the 250 $\mu$m resolution in Figure \ref{fig:ltir250}. The $\mathrm{L_{TIR}}(24+100)$ and $\mathrm{L_{TIR}}(24+70+100)$ show the least spatial variation when compared to the $\mathrm{L_{TIR}}(24+70+100+160+250)$ calibration and agree well (ratio$\approx$1) with this estimate. The remaining calibrations tend to predict higher or lower values in the overlap, particularly near SGMC345. 

The $\mathrm{L_{TIR}}(24+100)$ weights the 24 $\mu$m and 100 $\mu$m fluxes with coefficients of $2.453\pm0.085$ and $1.407\pm0.013$, while the $\mathrm{L_{TIR}}(24+70+100)$ calibration coefficients are $2.192\pm0.114$, $0.187\pm0.035$, and $1.314\pm0.016$ for the 24, 70, and 100 $\mu$m fluxes, respectively. The 24 $\mu$m and 100 $\mu$m fluxes appear to be similarly-weighted across these two calibrations, with the 70 $\mu$m flux being weighted relatively low for $\mathrm{L_{TIR}}(24+70+100)$. In comparison to the other calibrations which use the 70 $\mu$m flux, this is the lowest 70 $\mu$m coefficient. Because of the low/non-dependence of these calibrations on the 70 $\mu$m flux, these appear to reduce the (potential) effect of dust heating in the strong-starbursting environment of SGMC345. See \S \ref{sec:ltir} for the remainder of our discussion on the variation of different \citet{Galametz:2013} calibrations.

\citet{Galametz:2013} find the \textit{Herschel} 100$\mu$m band to be the best monochromatic estimate for L$_\mathrm{TIR}$ for their sample of galaxies (it is within $30\%$ of their SED-modelled L$_\mathrm{TIR}$ estimates). This calibration also shows little variation when compared with their modelled L$_\mathrm{TIR}$ (see Figure 7 in \citealt{Galametz:2013}) as a function of the 70/100 color.  The outliers of the 100 $\mu$m relationship were mainly strongly starbursting galaxies, NGC 1377 and NGC 5408, with SED peaks at lower IR wavelengths, $\sim$ 60 and 70 $\mu$m, respectively. \citet{Galametz:2013} find the 70 $\mu$m band tends to overestimate lower IR luminosity objects (L$_\mathrm{TIR} < 3\times10^8\ \mathrm{L}_\odot$) and suggest using the 70 $\mu$m band as an estimator for starbursting objects. Similarly, \citet{Galametz:2013} find the 160 $\mu$m calibration tends to underestimate L$_\mathrm{TIR}$ for hot objects, like starbursts or low-metallicity objects, and overestimate L$_\mathrm{TIR}$ for cooler objects. The 70 and 160 $\mu$m calibrations provide reasonable estimates of L$_\mathrm{TIR}$ to within $<50\%$, but \cite{Galametz:2013} suggest that the 70 and 160 $\mu$m calibrations are used with caution for hot or cold SEDs. \cite{Klaas:2010} plot IR SEDs of several clumps that are identified in $24-160\ \mu$m maps of the Antennae and find that the SED shape agrees well for most regions across this wavelength range, with peaks at $\sim100\ \mu$m. However, one clump in their study (which corresponds to the region in the overlap with SGMC345) shows a higher $24/70$ ratio ($\sim0.15$ vs. $\sim0.04-0.08$), with a hotter SED (peak at $\sim70\  \mu$m). With this in mind, we compare L$_\mathrm{TIR}$ from several multi-band L$_\mathrm{TIR}$ calibrations from \citet{Galametz:2013} in Figure \ref{fig:ltir250} and Table \ref{tab:ltir}.

\begin{table}
\centering
\caption{Total Infrared Luminosities from Different Calibrations}
\begin{tabular}{lccccc}\hline
Source	&	\multicolumn{1}{c}{24+70+100/70 Ratio}	&	\multicolumn{1}{c}{$\mathrm{L_{TIR}(70)}$}	&	\multicolumn{1}{c}{$\mathrm{L_{TIR}(70)}$} \\
	&	(at $6.8''$ res.)	&	\multicolumn{1}{c}{($10^9$ L$_\odot$ at $6.8''$ res.)}	& \multicolumn{2}{c}{($10^9$ L$_\odot$ at $5.5''$ res.)}	\\ \hline\hline 
NGC4038 & 0.95 & 8.1$\pm$0.4 & 8.3$\pm$0.4 \\
NGC4039 & 0.85 & 4.9$\pm$0.2 & 5.0$\pm$0.3 \\
NGC4038-2 & 0.88 & 0.67$\pm$0.03 & 0.66$\pm$0.03 \\
WArm-1 & 0.92 & 1.03$\pm$0.05 & 1.08$\pm$0.05 \\
WArm-3 & 0.83 & 2.8$\pm$0.1 & 2.9$\pm$0.1 \\
WArm-2 & 0.90 & 0.75$\pm$0.04 & 0.75$\pm$0.04 \\
WArm-4 & 0.86 & 1.87$\pm$0.09 & 2.0$\pm$0.1 \\
SGMC1 & 0.93 & 3.3$\pm$0.2 & 3.5$\pm$0.2 \\
SGMC2 & 0.88 & 5.3$\pm$0.3 & 5.3$\pm$0.3 \\
SGMC345 & 1.03 & 9.5$\pm$0.5 & 10.1$\pm$0.5 \\
Schirm-C6 & 0.81 & 2.5$\pm$0.1 & 2.5$\pm$0.1 \\
Schirm-C7 & 0.82 & 3.8$\pm$0.2 & 4.0$\pm$0.2 \\
Overlap-8 & 0.87 & 2.1$\pm$0.1 & 2.1$\pm$0.1 \\
Overlap-9 & 1.06 & 1.48$\pm$0.08 & 1.41$\pm$0.07 \\

\hline
\end{tabular}
\label{tab:ltir}
\end{table}

\section{Luminosity Ratios} \label{app:lumin_ratio}

We include line luminosity ratios in Table \ref{tab:ell_ratio} for the elliptical apertures. The table is divided into three sections: 1. The HCN and HCO$^+$ luminosity relative to CO. 2. The total infrared luminosity relative to CO, HCN, and HCO$^+$, and 3. Ratios of our three dense gas tracers, HCN, HCO$^+$, and HNC.


\begin{table*}[tb]
\centering
\begin{threeparttable}
\setlength{\tabcolsep}{3pt}
\caption{\sc Luminosity Ratios from Elliptical Apertures}
\label{tab:ell_ratio}
\begin{tabular}{lcc|ccc|ccc}\hline\hline
\multicolumn{1}{c}{Source}	&
\multicolumn{1}{c}{$\mathrm{L_{HCN}/L_{CO}}$}		&
\multicolumn{1}{c}{$\mathrm{L_{HCO^+}/L_{CO}}$}	&
\multicolumn{1}{|c}{$\mathrm{L_{TIR}/L_{CO}}$}		&
\multicolumn{1}{c}{$\mathrm{L_{TIR}/L_{HCN}}$}	&	
\multicolumn{1}{c|}{$\mathrm{L_{TIR}/L_{HCO^+}}$}	&	
\multicolumn{1}{c}{$\mathrm{L_{HCN}/L_{HCO^+}}$} 	&
\multicolumn{1}{c}{$\mathrm{L_{HNC}/L_{HCN}}$} 	&
\multicolumn{1}{c}{$\mathrm{L_{HNC}/L_{HCO^+}}$}	\\
\multicolumn{3}{c}{ }	&	\multicolumn{3}{|c|}{L$_\odot$ (K km s$^{-1}$ pc$^2)^{-1}$}	& \multicolumn{3}{c}{ }	\\ %
\\ [-1.0em] \hline
NGC4038 & 0.068$\pm$0.015 & 0.065$\pm$0.014 & 0.200$\pm$0.043 & 2.95$\pm$0.24 & 3.07$\pm$0.24 & 1.04$\pm$0.09 & 0.388$\pm$0.039 & 0.404$\pm$0.041 \\
NGC4039 & 0.050$\pm$0.012 & 0.075$\pm$0.017 & 0.234$\pm$0.053 & 4.67$\pm$0.48 & 3.14$\pm$0.29 & 0.673$\pm$0.081 & $<$0.436 & $<$0.293 \\
NGC4038-2 & 0.026$\pm$0.011 & 0.030$\pm$0.013 & 0.202$\pm$0.064 & 7.9$\pm$2.5 & 6.7$\pm$1.9 & 0.85$\pm$0.36 & $<$0.52 & $<$0.4 \\
WArm-1 & 0.0301$\pm$0.0094 & $<$0.03 & 0.244$\pm$0.062 & 8.1$\pm$1.6 & $>$7.4 & $>$0.91 & $<$0.79 & -- \\
WArm-2 & $<$0.017 & 0.0237$\pm$0.0079 & 0.161$\pm$0.041 & $>$9.5 & 6.8$\pm$1.5 & $<$0.71 & -- & $<$0.68 \\
WArm-3 & 0.047$\pm$0.015 & 0.038$\pm$0.012 & 0.70$\pm$0.18 & 14.9$\pm$2.9 & 18.3$\pm$3.5 & 1.23$\pm$0.33 & $<$0.79 & $<$0.63 \\
WArm-4 & 0.050$\pm$0.017 & 0.061$\pm$0.021 & 0.80$\pm$0.23 & 16.2$\pm$3.5 & 13.2$\pm$2.5 & 0.81$\pm$0.23 & 0.85$\pm$0.26 & 0.78$\pm$0.22 \\
SGMC1 & 0.026$\pm$0.007 & 0.062$\pm$0.015 & 0.193$\pm$0.044 & 7.4$\pm$1.2 & 3.14$\pm$0.32 & 0.422$\pm$0.073 & 0.45$\pm$0.12 & 0.192$\pm$0.044 \\
SGMC2 & 0.0225$\pm$0.0061 & 0.052$\pm$0.012 & 0.204$\pm$0.046 & 9.0$\pm$1.5 & 3.93$\pm$0.42 & 0.435$\pm$0.078 & 0.312$\pm$0.086 & 0.136$\pm$0.034 \\
SGMC345 & 0.0315$\pm$0.0081 & 0.061$\pm$0.015 & 0.57$\pm$0.13 & 18.0$\pm$2.6 & 9.31$\pm$0.96 & 0.516$\pm$0.082 & $<$0.352 & $<$0.182 \\
Schirm-C6 & 0.025$\pm$0.009 & 0.031$\pm$0.011 & 0.43$\pm$0.12 & 17.6$\pm$4.3 & 13.7$\pm$3.2 & 0.78$\pm$0.25 & $<$0.57 & $<$0.44 \\
Schirm-C7 & 0.0178$\pm$0.0057 & 0.0324$\pm$0.0098 & 0.50$\pm$0.12 & 27.8$\pm$6.0 & 15.3$\pm$2.9 & 0.55$\pm$0.15 & $<$0.49 & $<$0.271 \\
Overlap-8 & $<$0.0159 & 0.0216$\pm$0.0073 & 0.4$\pm$0.1 & $>$24.6 & 18.1$\pm$4.1 & $<$0.74 & -- & $<$0.61 \\
Overlap-9 & $<$0.028 & 0.104$\pm$0.039 & 0.90$\pm$0.29 & $>$32.0 & 8.6$\pm$1.8 & $<$0.270 & $>$1.01 & 0.273$\pm$0.093 \\

\hline
\end{tabular}
\begin{tablenotes}
\item {\sc Note. --} Luminosities measured from the elliptical apertures listed in Table \ref{tab:ell_ap}. All values are measured at the 100 $\mu$m resolution (6.8$''$). The absolute uncertainties are shown next to each ratio, except in the case of limits. We do not show ratios when both luminosity measurements are limits.
\end{tablenotes}
\end{threeparttable}
\end{table*}

\bibliographystyle{apj}
\bibliography{dense_gas}

\begin{thebibliography}{}
\expandafter\ifx\csname natexlab\endcsname\relax\def\natexlab#1{#1}\fi

\bibitem[{{Aniano} {et~al.}(2011){Aniano}, {Draine}, {Gordon}, \&
  {Sandstrom}}]{Aniano:2011}
{Aniano}, G., {Draine}, B.~T., {Gordon}, K.~D., \& {Sandstrom}, K. 2011, \pasp,
  123, 1218

\bibitem[{{Bastian} {et~al.}(2009){Bastian}, {Trancho}, {Konstantopoulos}, \&
  {Miller}}]{Bastian:2009}
{Bastian}, N., {Trancho}, G., {Konstantopoulos}, I.~S., \& {Miller}, B.~W.
  2009, \apj, 701, 607

\bibitem[{{Bendo} {et~al.}(2012{\natexlab{a}}){Bendo}, {Galliano}, \&
  {Madden}}]{Bendo:2012}
{Bendo}, G.~J., {Galliano}, F., \& {Madden}, S.~C. 2012{\natexlab{a}}, \mnras,
  423, 197

\bibitem[{{Bendo} {et~al.}(2012{\natexlab{b}}){Bendo}, {Boselli}, {Dariush},
  {Pohlen}, {Roussel}, {Sauvage}, {Smith}, {Wilson}, {Baes}, {Cooray},
  {Clements}, {Cortese}, {Foyle}, {Galametz}, {Gomez}, {Lebouteiller}, {Lu},
  {Madden}, {Mentuch}, {O'Halloran}, {Page}, {Remy}, {Schulz}, \&
  {Spinoglio}}]{Bendo:2012:VNGS}
{Bendo}, G.~J., {Boselli}, A., {Dariush}, A., {et~al.} 2012{\natexlab{b}},
  \mnras, 419, 1833

\bibitem[{{Bigiel} {et~al.}(2015){Bigiel}, {Leroy}, {Blitz}, {Bolatto}, {da
  Cunha}, {Rosolowsky}, {Sandstrom}, \& {Usero}}]{Bigiel:2015}
{Bigiel}, F., {Leroy}, A.~K., {Blitz}, L., {et~al.} 2015, \apj, 815, 103

\bibitem[{{Bigiel} {et~al.}(2016){Bigiel}, {Leroy}, {Jim{\'e}nez-Donaire},
  {Pety}, {Usero}, {Cormier}, {Bolatto}, {Garcia-Burillo}, {Colombo},
  {Gonz{\'a}lez-Garc{\'{\i}}a}, {Hughes}, {Kepley}, {Kramer}, {Sandstrom},
  {Schinnerer}, {Schruba}, {Schuster}, {Tomicic}, \&
  {Zschaechner}}]{Bigiel:2016}
{Bigiel}, F., {Leroy}, A.~K., {Jim{\'e}nez-Donaire}, M.~J., {et~al.} 2016,
  \apjl, 822, L26

\bibitem[{{Bolatto} {et~al.}(2013){Bolatto}, {Wolfire}, \&
  {Leroy}}]{Bolatto:2013}
{Bolatto}, A.~D., {Wolfire}, M., \& {Leroy}, A.~K. 2013, \araa, 51, 207

\bibitem[{{Booth} \& {Aalto}(1998)}]{Booth:1998}
{Booth}, R.~S., \& {Aalto}, S. 1998, The Molecular Astrophysics of Stars and
  Galaxies, edited by Thomas W.~Hartquist and David A.~Williams.~Clarendon
  Press, Oxford, 1998., p.437, 4, 437

\bibitem[{Bradley {et~al.}(2016)Bradley, Sipocz, Robitaille, Tollerud,
  Vinícius, Deil, Barbary, Günther, Cara, Droettboom, Bostroem, Bray,
  Bratholm, Pickering, Craig, Barentsen, Pascual, adonath, Greco, Kerzendorf,
  StuartLittlefair, Ferreira, D'Eugenio, \& Weaver}]{photutils}
Bradley, L., Sipocz, B., Robitaille, T., {et~al.} 2016, astropy/photutils:
  v0.3, doi:10.5281/zenodo.164986

\bibitem[{{Brandl} {et~al.}(2009){Brandl}, {Snijders}, {den Brok}, {Whelan},
  {Groves}, {van der Werf}, {Charmandaris}, {Smith}, {Armus}, {Kennicutt}, \&
  {Houck}}]{Brandl:2009}
{Brandl}, B.~R., {Snijders}, L., {den Brok}, M., {et~al.} 2009, \apj, 699, 1982

\bibitem[{{Chen} {et~al.}(2015){Chen}, {Gao}, {Braine}, \& {Gu}}]{Chen:2015}
{Chen}, H., {Gao}, Y., {Braine}, J., \& {Gu}, Q. 2015, \apj, 810, 140

\bibitem[{{Chown} {et~al.}(2018){Chown}, {Li}, {Li}, {Athanassoula}, {Wilson},
  {Lin}, {Mo}, {Parker}, \& {Xiao}}]{Chown:2018}
{Chown}, R., {Li}, C., {Li}, N., {et~al.} 2018, ArXiv e-prints,
  arXiv:1810.08624

\bibitem[{{Daddi} {et~al.}(2010){Daddi}, {Elbaz}, {Walter}, {Bournaud},
  {Salmi}, {Carilli}, {Dannerbauer}, {Dickinson}, {Monaco}, \&
  {Riechers}}]{Daddi:2010}
{Daddi}, E., {Elbaz}, D., {Walter}, F., {et~al.} 2010, \apjl, 714, L118

\bibitem[{{Federrath} \& {Klessen}(2012)}]{Federrath:2012}
{Federrath}, C., \& {Klessen}, R.~S. 2012, \apj, 761, 156

\bibitem[{{Galametz} {et~al.}(2013){Galametz}, {Kennicutt}, {Calzetti},
  {Aniano}, {Draine}, {Boquien}, {Brandl}, {Croxall}, {Dale}, {Engelbracht},
  {Gordon}, {Groves}, {Hao}, {Helou}, {Hinz}, {Hunt}, {Johnson}, {Li},
  {Murphy}, {Roussel}, {Sandstrom}, {Skibba}, \& {Tabatabaei}}]{Galametz:2013}
{Galametz}, M., {Kennicutt}, R.~C., {Calzetti}, D., {et~al.} 2013, \mnras, 431,
  1956

\bibitem[{{Gallagher} {et~al.}(2018){Gallagher}, {Leroy}, {Bigiel}, {Cormier},
  {Jim{\'e}nez-Donaire}, {Ostriker}, {Usero}, {Bolatto},
  {Garc{\'{\i}}a-Burillo}, {Hughes}, {Kepley}, {Krumholz}, {Meidt}, {Meier},
  {Murphy}, {Pety}, {Rosolowsky}, {Schinnerer}, {Schruba}, \&
  {Walter}}]{Gallagher:2018}
{Gallagher}, M.~J., {Leroy}, A.~K., {Bigiel}, F., {et~al.} 2018, ApJ, 858, 90

\bibitem[{{Gao} \& {Solomon}(2004{\natexlab{a}})}]{Gao:2004a}
{Gao}, Y., \& {Solomon}, P.~M. 2004{\natexlab{a}}, \apjs, 152, 63

\bibitem[{{Gao} \& {Solomon}(2004{\natexlab{b}})}]{Gao:2004b}
---. 2004{\natexlab{b}}, \apj, 606, 271

\bibitem[{{Gilbert} \& {Graham}(2007)}]{Gilbert:2007}
{Gilbert}, A.~M., \& {Graham}, J.~R. 2007, \apj, 668, 168

\bibitem[{{Gilbert} {et~al.}(2000){Gilbert}, {Graham}, {McLean}, {Becklin},
  {Figer}, {Larkin}, {Levenson}, {Teplitz}, \& {Wilcox}}]{Gilbert:2000}
{Gilbert}, A.~M., {Graham}, J.~R., {McLean}, I.~S., {et~al.} 2000, \apjl, 533,
  L57

\bibitem[{{Graci{\'a}-Carpio} {et~al.}(2008){Graci{\'a}-Carpio},
  {Garc{\'{\i}}a-Burillo}, {Planesas}, {Fuente}, \&
  {Usero}}]{Gracia-Carpio:2008}
{Graci{\'a}-Carpio}, J., {Garc{\'{\i}}a-Burillo}, S., {Planesas}, P., {Fuente},
  A., \& {Usero}, A. 2008, \aap, 479, 703

\bibitem[{{Griffin} {et~al.}(2010){Griffin}, {Abergel}, {Abreu}, {Ade},
  {Andr{\'e}}, {Augueres}, {Babbedge}, {Bae}, {Baillie}, {Baluteau}, {Barlow},
  {Bendo}, {Benielli}, {Bock}, {Bonhomme}, {Brisbin}, {Brockley-Blatt},
  {Caldwell}, {Cara}, {Castro-Rodriguez}, {Cerulli}, {Chanial}, {Chen},
  {Clark}, {Clements}, {Clerc}, {Coker}, {Communal}, {Conversi}, {Cox},
  {Crumb}, {Cunningham}, {Daly}, {Davis}, {de Antoni}, {Delderfield}, {Devin},
  {di Giorgio}, {Didschuns}, {Dohlen}, {Donati}, {Dowell}, {Dowell}, {Duband},
  {Dumaye}, {Emery}, {Ferlet}, {Ferrand}, {Fontignie}, {Fox}, {Franceschini},
  {Frerking}, {Fulton}, {Garcia}, {Gastaud}, {Gear}, {Glenn}, {Goizel},
  {Griffin}, {Grundy}, {Guest}, {Guillemet}, {Hargrave}, {Harwit}, {Hastings},
  {Hatziminaoglou}, {Herman}, {Hinde}, {Hristov}, {Huang}, {Imhof}, {Isaak},
  {Israelsson}, {Ivison}, {Jennings}, {Kiernan}, {King}, {Lange}, {Latter},
  {Laurent}, {Laurent}, {Leeks}, {Lellouch}, {Levenson}, {Li}, {Li},
  {Lilienthal}, {Lim}, {Liu}, {Lu}, {Madden}, {Mainetti}, {Marliani}, {McKay},
  {Mercier}, {Molinari}, {Morris}, {Moseley}, {Mulder}, {Mur}, {Naylor},
  {Nguyen}, {O'Halloran}, {Oliver}, {Olofsson}, {Olofsson}, {Orfei}, {Page},
  {Pain}, {Panuzzo}, {Papageorgiou}, {Parks}, {Parr-Burman}, {Pearce},
  {Pearson}, {P{\'e}rez-Fournon}, {Pinsard}, {Pisano}, {Podosek}, {Pohlen},
  {Polehampton}, {Pouliquen}, {Rigopoulou}, {Rizzo}, {Roseboom}, {Roussel},
  {Rowan-Robinson}, {Rownd}, {Saraceno}, {Sauvage}, {Savage}, {Savini},
  {Sawyer}, {Scharmberg}, {Schmitt}, {Schneider}, {Schulz}, {Schwartz},
  {Shafer}, {Shupe}, {Sibthorpe}, {Sidher}, {Smith}, {Smith}, {Smith},
  {Spencer}, {Stobie}, {Sudiwala}, {Sukhatme}, {Surace}, {Stevens}, {Swinyard},
  {Trichas}, {Tourette}, {Triou}, {Tseng}, {Tucker}, {Turner}, {Vaccari},
  {Valtchanov}, {Vigroux}, {Virique}, {Voellmer}, {Walker}, {Ward}, {Waskett},
  {Weilert}, {Wesson}, {White}, {Whitehouse}, {Wilson}, {Winter}, {Woodcraft},
  {Wright}, {Xu}, {Zavagno}, {Zemcov}, {Zhang}, \& {Zonca}}]{Griffin:2010}
{Griffin}, M.~J., {Abergel}, A., {Abreu}, A., {et~al.} 2010, \aap, 518, L3

\bibitem[{{Groves} {et~al.}(2008){Groves}, {Dopita}, {Sutherland}, {Kewley},
  {Fischera}, {Leitherer}, {Brandl}, \& {van Breugel}}]{Groves:2008}
{Groves}, B., {Dopita}, M.~A., {Sutherland}, R.~S., {et~al.} 2008, \apjs, 176,
  438

\bibitem[{{Hao} {et~al.}(2011){Hao}, {Kennicutt}, {Johnson}, {Calzetti},
  {Dale}, \& {Moustakas}}]{Hao:2011}
{Hao}, C.-N., {Kennicutt}, R.~C., {Johnson}, B.~D., {et~al.} 2011, \apj, 741,
  124

\bibitem[{{Helfer} \& {Blitz}(1997)}]{Helfer:1997}
{Helfer}, T.~T., \& {Blitz}, L. 1997, \apj, 478, 162

\bibitem[{{Herrera} \& {Boulanger}(2017)}]{Herrera:2017}
{Herrera}, C.~N., \& {Boulanger}, F. 2017, \aap, 600, A139

\bibitem[{{Hughes} {et~al.}(2013){Hughes}, {Meidt}, {Colombo}, {Schinnerer},
  {Pety}, {Leroy}, {Dobbs}, {Garc{\'{\i}}a-Burillo}, {Thompson}, {Dumas},
  {Schuster}, \& {Kramer}}]{Hughes:2013}
{Hughes}, A., {Meidt}, S.~E., {Colombo}, D., {et~al.} 2013, \apj, 779, 46

\bibitem[{Hunter(2007)}]{matplotlib}
Hunter, J.~D. 2007, Computing In Science \& Engineering, 9, 90

\bibitem[{{Jim{\'e}nez-Donaire} {et~al.}(2017){Jim{\'e}nez-Donaire}, {Bigiel},
  {Leroy}, {Cormier}, {Gallagher}, {Usero}, {Bolatto}, {Colombo},
  {Garc{\'{\i}}a-Burillo}, {Hughes}, {Kramer}, {Krumholz}, {Meier}, {Murphy},
  {Pety}, {Rosolowsky}, {Schinnerer}, {Schruba}, {Tomi{\v c}i{\'c}}, \&
  {Zschaechner}}]{JimenezDonaire:2017}
{Jim{\'e}nez-Donaire}, M.~J., {Bigiel}, F., {Leroy}, A.~K., {et~al.} 2017,
  \mnras, 466, 49

\bibitem[{{Johnson} {et~al.}(2015){Johnson}, {Leroy}, {Indebetouw}, {Brogan},
  {Whitmore}, {Hibbard}, {Sheth}, \& {Evans}}]{Johnson:2015}
{Johnson}, K.~E., {Leroy}, A.~K., {Indebetouw}, R., {et~al.} 2015, \apj, 806,
  35

\bibitem[{{Jones} {et~al.}(2012){Jones}, {Burton}, {Cunningham},
  {Requena-Torres}, {Menten}, {Schilke}, {Belloche}, {Leurini},
  {Mart{\'{\i}}n-Pintado}, {Ott}, \& {Walsh}}]{Jones:2012}
{Jones}, P.~A., {Burton}, M.~G., {Cunningham}, M.~R., {et~al.} 2012, MNRAS,
  419, 2961

\bibitem[{{Karl} {et~al.}(2010){Karl}, {Naab}, {Johansson}, {Kotarba}, {Boily},
  {Renaud}, \& {Theis}}]{Karl:2010}
{Karl}, S.~J., {Naab}, T., {Johansson}, P.~H., {et~al.} 2010, \apjl, 715, L88

\bibitem[{{Kauffmann} {et~al.}(2017{\natexlab{a}}){Kauffmann}, {Goldsmith},
  {Melnick}, {Tolls}, {Guzman}, \& {Menten}}]{Kauffmann:2017c}
{Kauffmann}, J., {Goldsmith}, P.~F., {Melnick}, G., {et~al.}
  2017{\natexlab{a}}, \aap, 605, L5

\bibitem[{{Kauffmann} {et~al.}(2017{\natexlab{b}}){Kauffmann}, {Pillai},
  {Zhang}, {Menten}, {Goldsmith}, {Lu}, \& {Guzm{\'a}n}}]{Kauffmann:2017a}
{Kauffmann}, J., {Pillai}, T., {Zhang}, Q., {et~al.} 2017{\natexlab{b}}, \aap,
  603, A89

\bibitem[{{Kauffmann} {et~al.}(2017{\natexlab{c}}){Kauffmann}, {Pillai},
  {Zhang}, {Menten}, {Goldsmith}, {Lu}, {Guzm{\'a}n}, \&
  {Schmiedeke}}]{Kauffmann:2017b}
---. 2017{\natexlab{c}}, \aap, 603, A90

\bibitem[{{Kelly}(2007)}]{Kelly:2007}
{Kelly}, B.~C. 2007, \apj, 665, 1489

\bibitem[{{Kennicutt} \& {Evans}(2012)}]{Kennicutt:2012}
{Kennicutt}, R.~C., \& {Evans}, N.~J. 2012, \araa, 50, 531

\bibitem[{{Kennicutt}(1998)}]{Kennicutt:1998}
{Kennicutt}, Jr., R.~C. 1998, \araa, 36, 189

\bibitem[{{Kepley} {et~al.}(2014){Kepley}, {Leroy}, {Frayer}, {Usero},
  {Marvil}, \& {Walter}}]{Kepley:2014}
{Kepley}, A.~A., {Leroy}, A.~K., {Frayer}, D., {et~al.} 2014, ApJL, 780, L13

\bibitem[{{Klaas} {et~al.}(2010){Klaas}, {Nielbock}, {Haas}, {Krause}, \&
  {Schreiber}}]{Klaas:2010}
{Klaas}, U., {Nielbock}, M., {Haas}, M., {Krause}, O., \& {Schreiber}, J. 2010,
  \aap, 518, L44

\bibitem[{{Kruijssen} {et~al.}(2014){Kruijssen}, {Longmore}, {Elmegreen},
  {Murray}, {Bally}, {Testi}, \& {Kennicutt}}]{Kruijssen:2014}
{Kruijssen}, J.~M.~D., {Longmore}, S.~N., {Elmegreen}, B.~G., {et~al.} 2014,
  \mnras, 440, 3370

\bibitem[{{Krumholz} \& {McKee}(2005)}]{KrumholzMcKee:2005}
{Krumholz}, M.~R., \& {McKee}, C.~F. 2005, \apj, 630, 250

\bibitem[{{Lada} {et~al.}(2012){Lada}, {Forbrich}, {Lombardi}, \&
  {Alves}}]{Lada:2012}
{Lada}, C.~J., {Forbrich}, J., {Lombardi}, M., \& {Alves}, J.~F. 2012, \apj,
  745, 190

\bibitem[{{Lada} {et~al.}(1991{\natexlab{a}}){Lada}, {Bally}, \&
  {Stark}}]{Lada:1991a}
{Lada}, E.~A., {Bally}, J., \& {Stark}, A.~A. 1991{\natexlab{a}}, \apj, 368,
  432

\bibitem[{{Lada} {et~al.}(1991{\natexlab{b}}){Lada}, {Depoy}, {Evans}, \&
  {Gatley}}]{Lada:1991b}
{Lada}, E.~A., {Depoy}, D.~L., {Evans}, II, N.~J., \& {Gatley}, I.
  1991{\natexlab{b}}, \apj, 371, 171

\bibitem[{{Leroy} {et~al.}(2016){Leroy}, {Hughes}, {Schruba}, {Rosolowsky},
  {Blanc}, {Bolatto}, {Colombo}, {Escala}, {Kramer}, {Kruijssen}, {Meidt},
  {Pety}, {Querejeta}, {Sandstrom}, {Schinnerer}, {Sliwa}, \&
  {Usero}}]{Leroy:2016}
{Leroy}, A.~K., {Hughes}, A., {Schruba}, A., {et~al.} 2016, \apj, 831, 16

\bibitem[{{Leroy} {et~al.}(2017){Leroy}, {Usero}, {Schruba}, {Bigiel},
  {Kruijssen}, {Kepley}, {Blanc}, {Bolatto}, {Cormier}, {Gallagher}, {Hughes},
  {Jim{\'e}nez-Donaire}, {Rosolowsky}, \& {Schinnerer}}]{Leroy:2017}
{Leroy}, A.~K., {Usero}, A., {Schruba}, A., {et~al.} 2017, \apj, 835, 217

\bibitem[{{Liu} {et~al.}(2015){Liu}, {Gao}, \& {Greve}}]{Liu:2015}
{Liu}, L., {Gao}, Y., \& {Greve}, T.~R. 2015, \apj, 805, 31

\bibitem[{{Loenen} {et~al.}(2008){Loenen}, {Spaans}, {Baan}, \&
  {Meijerink}}]{Loenen:2008}
{Loenen}, A.~F., {Spaans}, M., {Baan}, W.~A., \& {Meijerink}, R. 2008, \aap,
  488, L5

\bibitem[{{Mao} {et~al.}(2000){Mao}, {Henkel}, {Schulz}, {Zielinsky},
  {Mauersberger}, {St{\"o}rzer}, {Wilson}, \& {Gensheimer}}]{Mao:2000}
{Mao}, R.~Q., {Henkel}, C., {Schulz}, A., {et~al.} 2000, \aap, 358, 433

\bibitem[{Mckinney(2010)}]{pandas}
Mckinney, W. 2010

\bibitem[{{McMullin} {et~al.}(2007){McMullin}, {Waters}, {Schiebel}, {Young},
  \& {Golap}}]{McMullin:2007}
{McMullin}, J.~P., {Waters}, B., {Schiebel}, D., {Young}, W., \& {Golap}, K.
  2007, in Astronomical Society of the Pacific Conference Series, Vol. 376,
  Astronomical Data Analysis Software and Systems XVI, ed. R.~A. {Shaw},
  F.~{Hill}, \& D.~J. {Bell}, 127

\bibitem[{{Meijerink} \& {Spaans}(2005)}]{Meijerink:2005}
{Meijerink}, R., \& {Spaans}, M. 2005, \aap, 436, 397

\bibitem[{{Meijerink} {et~al.}(2007){Meijerink}, {Spaans}, \&
  {Israel}}]{Meijerink:2007}
{Meijerink}, R., {Spaans}, M., \& {Israel}, F.~P. 2007, \aap, 461, 793

\bibitem[{{Mengel} {et~al.}(2005){Mengel}, {Lehnert}, {Thatte}, \&
  {Genzel}}]{Mengel:2005}
{Mengel}, S., {Lehnert}, M.~D., {Thatte}, N., \& {Genzel}, R. 2005, \aap, 443,
  41

\bibitem[{{Mengel} {et~al.}(2001){Mengel}, {Lehnert}, {Thatte},
  {Tacconi-Garman}, \& {Genzel}}]{Mengel:2001}
{Mengel}, S., {Lehnert}, M.~D., {Thatte}, N., {Tacconi-Garman}, L.~E., \&
  {Genzel}, R. 2001, \apj, 550, 280

\bibitem[{{Mihos} \& {Hernquist}(1996)}]{Mihos:1996}
{Mihos}, J.~C., \& {Hernquist}, L. 1996, \apj, 464, 641

\bibitem[{{Murphy} {et~al.}(2011){Murphy}, {Condon}, {Schinnerer}, {Kennicutt},
  {Calzetti}, {Armus}, {Helou}, {Turner}, {Aniano}, {Beir{\~a}o}, {Bolatto},
  {Brandl}, {Croxall}, {Dale}, {Donovan Meyer}, {Draine}, {Engelbracht},
  {Hunt}, {Hao}, {Koda}, {Roussel}, {Skibba}, \& {Smith}}]{Murphy:2011}
{Murphy}, E.~J., {Condon}, J.~J., {Schinnerer}, E., {et~al.} 2011, \apj, 737,
  67

\bibitem[{{Narayanan} {et~al.}(2011){Narayanan}, {Krumholz}, {Ostriker}, \&
  {Hernquist}}]{Narayanan:2011}
{Narayanan}, D., {Krumholz}, M., {Ostriker}, E.~C., \& {Hernquist}, L. 2011,
  \mnras, 418, 664

\bibitem[{{Neff} \& {Ulvestad}(2000)}]{Neff:2000}
{Neff}, S.~G., \& {Ulvestad}, J.~S. 2000, \aj, 120, 670

\bibitem[{{Onus} {et~al.}(2018){Onus}, {Krumholz}, \& {Federrath}}]{Onus:2018}
{Onus}, A., {Krumholz}, M.~R., \& {Federrath}, C. 2018, ArXiv e-prints,
  arXiv:1801.09952

\bibitem[{{Padoan} \& {Nordlund}(2011)}]{Padoan:2011}
{Padoan}, P., \& {Nordlund}, {\AA}. 2011, \apj, 730, 40

\bibitem[{{Papadopoulos}(2007)}]{Papadopoulos:2007}
{Papadopoulos}, P.~P. 2007, \apj, 656, 792

\bibitem[{{Papadopoulos} {et~al.}(2014){Papadopoulos}, {Zhang}, {Xilouris},
  {Weiss}, {van der Werf}, {Israel}, {Greve}, {Isaak}, \&
  {Gao}}]{Papadopoulos:2014}
{Papadopoulos}, P.~P., {Zhang}, Z.-Y., {Xilouris}, E.~M., {et~al.} 2014, \apj,
  788, 153

\bibitem[{{Pilbratt} {et~al.}(2010){Pilbratt}, {Riedinger}, {Passvogel},
  {Crone}, {Doyle}, {Gageur}, {Heras}, {Jewell}, {Metcalfe}, {Ott}, \&
  {Schmidt}}]{Pilbratt:2010}
{Pilbratt}, G.~L., {Riedinger}, J.~R., {Passvogel}, T., {et~al.} 2010, \aap,
  518, L1

\bibitem[{{Poglitsch} {et~al.}(2010){Poglitsch}, {Waelkens}, {Geis},
  {Feuchtgruber}, {Vandenbussche}, {Rodriguez}, {Krause}, {Renotte}, {van
  Hoof}, {Saraceno}, {Cepa}, {Kerschbaum}, {Agn{\`e}se}, {Ali}, {Altieri},
  {Andreani}, {Augueres}, {Balog}, {Barl}, {Bauer}, {Belbachir}, {Benedettini},
  {Billot}, {Boulade}, {Bischof}, {Blommaert}, {Callut}, {Cara}, {Cerulli},
  {Cesarsky}, {Contursi}, {Creten}, {De Meester}, {Doublier}, {Doumayrou},
  {Duband}, {Exter}, {Genzel}, {Gillis}, {Gr{\"o}zinger}, {Henning},
  {Herreros}, {Huygen}, {Inguscio}, {Jakob}, {Jamar}, {Jean}, {de Jong},
  {Katterloher}, {Kiss}, {Klaas}, {Lemke}, {Lutz}, {Madden}, {Marquet},
  {Martignac}, {Mazy}, {Merken}, {Montfort}, {Morbidelli}, {M{\"u}ller},
  {Nielbock}, {Okumura}, {Orfei}, {Ottensamer}, {Pezzuto}, {Popesso},
  {Putzeys}, {Regibo}, {Reveret}, {Royer}, {Sauvage}, {Schreiber}, {Stegmaier},
  {Schmitt}, {Schubert}, {Sturm}, {Thiel}, {Tofani}, {Vavrek}, {Wetzstein},
  {Wieprecht}, \& {Wiezorrek}}]{Poglitsch:2010}
{Poglitsch}, A., {Waelkens}, C., {Geis}, N., {et~al.} 2010, \aap, 518, L2

\bibitem[{{Rathborne} {et~al.}(2014){Rathborne}, {Longmore}, {Jackson},
  {Kruijssen}, {Alves}, {Bally}, {Bastian}, {Contreras}, {Foster}, {Garay},
  {Testi}, \& {Walsh}}]{Rathborne:2014}
{Rathborne}, J.~M., {Longmore}, S.~N., {Jackson}, J.~M., {et~al.} 2014, \apjl,
  795, L25

\bibitem[{{Reid} {et~al.}(2014){Reid}, {McClintock}, {Steiner}, {Steeghs},
  {Remillard}, {Dhawan}, \& {Narayan}}]{Reid:2014}
{Reid}, M.~J., {McClintock}, J.~E., {Steiner}, J.~F., {et~al.} 2014, ApJ, 796,
  2

\bibitem[{{Renaud} {et~al.}(2015){Renaud}, {Bournaud}, \& {Duc}}]{Renaud:2015}
{Renaud}, F., {Bournaud}, F., \& {Duc}, P.-A. 2015, \mnras, 446, 2038

\bibitem[{{Rieke} {et~al.}(2004){Rieke}, {Young}, {Engelbracht}, {Kelly},
  {Low}, {Haller}, {Beeman}, {Gordon}, {Stansberry}, {Misselt}, {Cadien},
  {Morrison}, {Rivlis}, {Latter}, {Noriega-Crespo}, {Padgett}, {Stapelfeldt},
  {Hines}, {Egami}, {Muzerolle}, {Alonso-Herrero}, {Blaylock}, {Dole}, {Hinz},
  {Le Floc'h}, {Papovich}, {P{\'e}rez-Gonz{\'a}lez}, {Smith}, {Su}, {Bennett},
  {Frayer}, {Henderson}, {Lu}, {Masci}, {Pesenson}, {Rebull}, {Rho}, {Keene},
  {Stolovy}, {Wachter}, {Wheaton}, {Werner}, \& {Richards}}]{Rieke:2004:MIPS}
{Rieke}, G.~H., {Young}, E.~T., {Engelbracht}, C.~W., {et~al.} 2004, \apjs,
  154, 25

\bibitem[{{Rosolowsky} \& {Leroy}(2006)}]{Rosolowsky:2006}
{Rosolowsky}, E., \& {Leroy}, A. 2006, \pasp, 118, 590

\bibitem[{{Rubin} {et~al.}(1970){Rubin}, {Ford}, \& {D'Odorico}}]{Rubin:1970}
{Rubin}, V.~C., {Ford}, Jr., W.~K., \& {D'Odorico}, S. 1970, \apj, 160, 801

\bibitem[{{Sanders} \& {Mirabel}(1996)}]{Sanders:1996}
{Sanders}, D.~B., \& {Mirabel}, I.~F. 1996, ARAA, 34, 749

\bibitem[{{Sandstrom} {et~al.}(2013){Sandstrom}, {Leroy}, {Walter}, {Bolatto},
  {Croxall}, {Draine}, {Wilson}, {Wolfire}, {Calzetti}, {Kennicutt}, {Aniano},
  {Donovan Meyer}, {Usero}, {Bigiel}, {Brinks}, {de Blok}, {Crocker}, {Dale},
  {Engelbracht}, {Galametz}, {Groves}, {Hunt}, {Koda}, {Kreckel}, {Linz},
  {Meidt}, {Pellegrini}, {Rix}, {Roussel}, {Schinnerer}, {Schruba}, {Schuster},
  {Skibba}, {van der Laan}, {Appleton}, {Armus}, {Brandl}, {Gordon}, {Hinz},
  {Krause}, {Montiel}, {Sauvage}, {Schmiedeke}, {Smith}, \&
  {Vigroux}}]{Sandstrom:2013}
{Sandstrom}, K.~M., {Leroy}, A.~K., {Walter}, F., {et~al.} 2013, \apj, 777, 5

\bibitem[{{Schilke} {et~al.}(1992){Schilke}, {Walmsley}, {Pineau Des Forets},
  {Roueff}, {Flower}, \& {Guilloteau}}]{Schilke:1992}
{Schilke}, P., {Walmsley}, C.~M., {Pineau Des Forets}, G., {et~al.} 1992, \aap,
  256, 595

\bibitem[{{Schirm} {et~al.}(2016){Schirm}, {Wilson}, {Madden}, \&
  {Clements}}]{Schirm:2016}
{Schirm}, M.~R.~P., {Wilson}, C.~D., {Madden}, S.~C., \& {Clements}, D.~L.
  2016, \apj, 823, 87

\bibitem[{{Schirm} {et~al.}(2014){Schirm}, {Wilson}, {Parkin}, {Kamenetzky},
  {Glenn}, {Rangwala}, {Spinoglio}, {Pereira-Santaella}, {Baes}, {Barlow},
  {Clements}, {Cooray}, {De Looze}, {Karczewski}, {Madden}, {R{\'e}my-Ruyer},
  \& {Wu}}]{Schirm:2014}
{Schirm}, M.~R.~P., {Wilson}, C.~D., {Parkin}, T.~J., {et~al.} 2014, \apj, 781,
  101

\bibitem[{{Schweizer} {et~al.}(2008){Schweizer}, {Burns}, {Madore}, {Mager},
  {Phillips}, {Freedman}, {Boldt}, {Contreras}, {Folatelli}, {Gonz{\'a}lez},
  {Hamuy}, {Krzeminski}, {Morrell}, {Persson}, {Roth}, \&
  {Stritzinger}}]{Schweizer:2008}
{Schweizer}, F., {Burns}, C.~R., {Madore}, B.~F., {et~al.} 2008, \aj, 136, 1482

\bibitem[{{Shimajiri} {et~al.}(2017){Shimajiri}, {Andr{\'e}}, {Braine},
  {K{\"o}nyves}, {Schneider}, {Bontemps}, {Ladjelate}, {Roy}, {Gao}, \&
  {Chen}}]{Shimajiri:2017}
{Shimajiri}, Y., {Andr{\'e}}, P., {Braine}, J., {et~al.} 2017, AAP, 604, A74

\bibitem[{{Shirley}(2015)}]{Shirley:2015}
{Shirley}, Y.~L. 2015, \pasp, 127, 299

\bibitem[{{Snijders} {et~al.}(2007){Snijders}, {Kewley}, \& {van der
  Werf}}]{Snijders:2007}
{Snijders}, L., {Kewley}, L.~J., \& {van der Werf}, P.~P. 2007, \apj, 669, 269

\bibitem[{{Stephens} {et~al.}(2016){Stephens}, {Jackson}, {Whitaker},
  {Contreras}, {Guzm{\'a}n}, {Sanhueza}, {Foster}, \&
  {Rathborne}}]{Stephens:2016}
{Stephens}, I.~W., {Jackson}, J.~M., {Whitaker}, J.~S., {et~al.} 2016, \apj,
  824, 29

\bibitem[{{Tacconi} {et~al.}(2018){Tacconi}, {Genzel}, {Saintonge}, {Combes},
  {Garc{\'{\i}}a-Burillo}, {Neri}, {Bolatto}, {Contini}, {F{\"o}rster
  Schreiber}, {Lilly}, {Lutz}, {Wuyts}, {Accurso}, {Boissier}, {Boone},
  {Bouch{\'e}}, {Bournaud}, {Burkert}, {Carollo}, {Cooper}, {Cox}, {Feruglio},
  {Freundlich}, {Herrera-Camus}, {Juneau}, {Lippa}, {Naab}, {Renzini},
  {Salome}, {Sternberg}, {Tadaki}, {{\"U}bler}, {Walter}, {Weiner}, \&
  {Weiss}}]{Tacconi:2018}
{Tacconi}, L.~J., {Genzel}, R., {Saintonge}, A., {et~al.} 2018, \apj, 853, 179

\bibitem[{{Talbi} {et~al.}(1996){Talbi}, {Ellinger}, \& {Herbst}}]{Talbi:1996}
{Talbi}, D., {Ellinger}, Y., \& {Herbst}, E. 1996, \aap, 314, 688

\bibitem[{{The Astropy Collaboration} {et~al.}(2018){The Astropy
  Collaboration}, {Price-Whelan}, {Sip{\H o}cz}, {G{\"u}nther}, {Lim},
  {Crawford}, {Conseil}, {Shupe}, {Craig}, {Dencheva}, {Ginsburg},
  {VanderPlas}, {Bradley}, {P{\'e}rez-Su{\'a}rez}, {de Val-Borro}, {Aldcroft},
  {Cruz}, {Robitaille}, {Tollerud}, {Ardelean}, {Babej}, {Bachetti}, {Bakanov},
  {Bamford}, {Barentsen}, {Barmby}, {Baumbach}, {Berry}, {Biscani}, {Boquien},
  {Bostroem}, {Bouma}, {Brammer}, {Bray}, {Breytenbach}, {Buddelmeijer},
  {Burke}, {Calderone}, {Cano Rodr{\'{\i}}guez}, {Cara}, {Cardoso},
  {Cheedella}, {Copin}, {Crichton}, {D{\'A}vella}, {Deil}, {Depagne},
  {Dietrich}, {Donath}, {Droettboom}, {Earl}, {Erben}, {Fabbro}, {Ferreira},
  {Finethy}, {Fox}, {Garrison}, {Gibbons}, {Goldstein}, {Gommers}, {Greco},
  {Greenfield}, {Groener}, {Grollier}, {Hagen}, {Hirst}, {Homeier}, {Horton},
  {Hosseinzadeh}, {Hu}, {Hunkeler}, {Ivezi{\'c}}, {Jain}, {Jenness}, {Kanarek},
  {Kendrew}, {Kern}, {Kerzendorf}, {Khvalko}, {King}, {Kirkby}, {Kulkarni},
  {Kumar}, {Lee}, {Lenz}, {Littlefair}, {Ma}, {Macleod}, {Mastropietro},
  {McCully}, {Montagnac}, {Morris}, {Mueller}, {Mumford}, {Muna}, {Murphy},
  {Nelson}, {Nguyen}, {Ninan}, {N{\"o}the}, {Ogaz}, {Oh}, {Parejko}, {Parley},
  {Pascual}, {Patil}, {Patil}, {Plunkett}, {Prochaska}, {Rastogi}, {Reddy
  Janga}, {Sabater}, {Sakurikar}, {Seifert}, {Sherbert}, {Sherwood-Taylor},
  {Shih}, {Sick}, {Silbiger}, {Singanamalla}, {Singer}, {Sladen}, {Sooley},
  {Sornarajah}, {Streicher}, {Teuben}, {Thomas}, {Tremblay}, {Turner},
  {Terr{\'o}n}, {van Kerkwijk}, {de la Vega}, {Watkins}, {Weaver}, {Whitmore},
  {Woillez}, \& {Zabalza}}]{astropy}
{The Astropy Collaboration}, {Price-Whelan}, A.~M., {Sip{\H o}cz}, B.~M.,
  {et~al.} 2018, ArXiv e-prints, arXiv:1801.02634

\bibitem[{{Toomre}(1977)}]{Toomre:1977}
{Toomre}, A. 1977, in Evolution of Galaxies and Stellar Populations, ed. B.~M.
  {Tinsley} \& R.~B.~G. {Larson}, D.~Campbell, 401

\bibitem[{{Usero} {et~al.}(2015){Usero}, {Leroy}, {Walter}, {Schruba},
  {Garc{\'{\i}}a-Burillo}, {Sandstrom}, {Bigiel}, {Brinks}, {Kramer},
  {Rosolowsky}, {Schuster}, \& {de Blok}}]{Usero:2015}
{Usero}, A., {Leroy}, A.~K., {Walter}, F., {et~al.} 2015, \aj, 150, 115

\bibitem[{{Utomo} {et~al.}(2017){Utomo}, {Bolatto}, {Wong}, {Ostriker},
  {Blitz}, {Sanchez}, {Colombo}, {Leroy}, {Cao}, {Dannerbauer},
  {Garcia-Benito}, {Husemann}, {Kalinova}, {Levy}, {Mast}, {Rosolowsky}, \&
  {Vogel}}]{Utomo:2017}
{Utomo}, D., {Bolatto}, A.~D., {Wong}, T., {et~al.} 2017, \apj, 849, 26

\bibitem[{{van der Tak} {et~al.}(2007){van der Tak}, {Black}, {Sch{\"o}ier},
  {Jansen}, \& {van Dishoeck}}]{vanderTak:2007}
{van der Tak}, F.~F.~S., {Black}, J.~H., {Sch{\"o}ier}, F.~L., {Jansen}, D.~J.,
  \& {van Dishoeck}, E.~F. 2007, \aap, 468, 627

\bibitem[{{Van Der Walt} {et~al.}(2011){Van Der Walt}, {Colbert}, \&
  {Varoquaux}}]{numpy}
{Van Der Walt}, S., {Colbert}, S.~C., \& {Varoquaux}, G. 2011, ArXiv e-prints,
  arXiv:1102.1523

\bibitem[{{Werner} {et~al.}(2004){Werner}, {Roellig}, {Low}, {Rieke}, {Rieke},
  {Hoffmann}, {Young}, {Houck}, {Brandl}, {Fazio}, {Hora}, {Gehrz}, {Helou},
  {Soifer}, {Stauffer}, {Keene}, {Eisenhardt}, {Gallagher}, {Gautier}, {Irace},
  {Lawrence}, {Simmons}, {Van Cleve}, {Jura}, {Wright}, \&
  {Cruikshank}}]{Werner:2004:Spitzer}
{Werner}, M.~W., {Roellig}, T.~L., {Low}, F.~J., {et~al.} 2004, \apjs, 154, 1

\bibitem[{{Whitmore} \& {Schweizer}(1995)}]{Whitmore:1995}
{Whitmore}, B.~C., \& {Schweizer}, F. 1995, \aj, 109, 960

\bibitem[{{Whitmore} \& {Zhang}(2002)}]{Whitmore:2002}
{Whitmore}, B.~C., \& {Zhang}, Q. 2002, \aj, 124, 1418

\bibitem[{{Whitmore} {et~al.}(1999){Whitmore}, {Zhang}, {Leitherer}, {Fall},
  {Schweizer}, \& {Miller}}]{Whitmore:1999}
{Whitmore}, B.~C., {Zhang}, Q., {Leitherer}, C., {et~al.} 1999, \aj, 118, 1551

\bibitem[{{Whitmore} {et~al.}(2010){Whitmore}, {Chandar}, {Schweizer},
  {Rothberg}, {Leitherer}, {Rieke}, {Rieke}, {Blair}, {Mengel}, \&
  {Alonso-Herrero}}]{Whitmore:2010}
{Whitmore}, B.~C., {Chandar}, R., {Schweizer}, F., {et~al.} 2010, \aj, 140, 75

\bibitem[{{Whitmore} {et~al.}(2014){Whitmore}, {Brogan}, {Chandar}, {Evans},
  {Hibbard}, {Johnson}, {Leroy}, {Privon}, {Remijan}, \&
  {Sheth}}]{Whitmore:2014}
{Whitmore}, B.~C., {Brogan}, C., {Chandar}, R., {et~al.} 2014, \apj, 795, 156

\bibitem[{{Wilson} {et~al.}(2000){Wilson}, {Scoville}, {Madden}, \&
  {Charmandaris}}]{Wilson:2000}
{Wilson}, C.~D., {Scoville}, N., {Madden}, S.~C., \& {Charmandaris}, V. 2000,
  \apj, 542, 120

\bibitem[{{Wilson} {et~al.}(2003){Wilson}, {Scoville}, {Madden}, \&
  {Charmandaris}}]{Wilson:2003}
---. 2003, \apj, 599, 1049

\bibitem[{{Wilson} {et~al.}(2008){Wilson}, {Petitpas}, {Iono}, {Baker}, {Peck},
  {Krips}, {Warren}, {Golding}, {Atkinson}, {Armus}, {Cox}, {Ho}, {Juvela},
  {Matsushita}, {Mihos}, {Pihlstrom}, \& {Yun}}]{Wilson:2008}
{Wilson}, C.~D., {Petitpas}, G.~R., {Iono}, D., {et~al.} 2008, \apjs, 178, 189

\bibitem[{{Wu} {et~al.}(2005){Wu}, {Evans}, {Gao}, {Solomon}, {Shirley}, \&
  {Vanden Bout}}]{Wu:2005}
{Wu}, J., {Evans}, II, N.~J., {Gao}, Y., {et~al.} 2005, \apjl, 635, L173

\bibitem[{{Wu} {et~al.}(2010){Wu}, {Evans}, {Shirley}, \& {Knez}}]{Wu:2010}
{Wu}, J., {Evans}, II, N.~J., {Shirley}, Y.~L., \& {Knez}, C. 2010, \apjs, 188,
  313

\bibitem[{{Zezas} {et~al.}(2002){Zezas}, {Fabbiano}, {Rots}, \&
  {Murray}}]{Zezas:2002}
{Zezas}, A., {Fabbiano}, G., {Rots}, A.~H., \& {Murray}, S.~S. 2002, \apj, 577,
  710

\bibitem[{{Zhu} {et~al.}(2003){Zhu}, {Seaquist}, \& {Kuno}}]{Zhu:2003}
{Zhu}, M., {Seaquist}, E.~R., \& {Kuno}, N. 2003, \apj, 588, 243

\end{thebibliography}

\end{document}